\documentclass[notitlepage,aps,pra,reprint,twocolumn,superscriptaddress]{revtex4-2}
\usepackage{amsmath,amssymb,braket,comment,float,graphicx,physics,ulem,wrapfig}
\usepackage[export]{adjustbox}
\usepackage[colorlinks, allcolors=blue]{hyperref}
\usepackage[all]{hypcap}
\usepackage{bm}
\usepackage{verbatim}
\usepackage{overpic}
\usepackage{orcidlink}

\newcommand{\ie}{\textit{i.e.} }
\newcommand{\eg}{\textit{e.g.} }

\DeclareMathOperator{\diag}{diag}

\begin{document}
\title{Fokker-Planck approach to thermal fluctuations in antiferromagnetic systems}

\author{E. Martello~\orcidlink{0000-0002-9041-4949}}
\affiliation{Department of Physics and Astronomy ``Ettore Majorana'', University of Catania, Catania, Italy}

\author{G. A. Falci~\orcidlink{0000-0001-5842-2677}}
\affiliation{Department of Physics and Astronomy ``Ettore Majorana'', University of Catania, Catania, Italy}
\affiliation{INFN, Catania Sec., I-95123, Catania, Italy}
\author{E. Paladino~\orcidlink{0000-0002-9929-3768}}
\affiliation{Department of Physics and Astronomy ``Ettore Majorana'', University of Catania, Catania, Italy}
\affiliation{INFN, Catania Sec., I-95123, Catania, Italy}

\author{F. M. D. Pellegrino~\orcidlink{0000-0001-5425-1292}}
\affiliation{Department of Physics and Astronomy ``Ettore Majorana'', University of Catania, Catania, Italy}
\affiliation{INFN, Catania Sec., I-95123, Catania, Italy}
\date{\today}
\begin{abstract}
    We develop a Fokker-Planck approach to describe the dynamics of staggered magnetization and thermal fluctuations in a two-dimensional antiferromagnetic system with uniaxial anisotropy. 
    Beginning with a classical model for the antiferromagnetic system, we incorporate a Landau-Lifshitz-Gilbert equation augmented by Langevin fields to account for thermal fluctuations, and we derive the Fokker-Planck equation governing the probability distribution function of the spin configuration.
    Employing the mean-field approximation, we derive the equations of motion for the spin polarization and the two-time spin-spin correlation functions. The methodology is applied to the study of spin-wave dynamics and to the formulation of a phenomenological model for resistance fluctuations in two-dimensional antiferromagnetic semiconductors.
\end{abstract}

\maketitle

\section{Introduction}

Two-dimensional (2D) magnetic materials have recently emerged as a frontier in condensed matter physics, offering a unique platform for exploring fundamental spin physics and developing next-generation spintronics~\cite{du2020berry,zhang2020gate,wang2023intrinsic,pancholi2025journey,liu2020robust}. 
The experimental ability to isolate atomically thin layers in van der Waals crystals has enabled direct access to intrinsic magnetic order in materials where the reduced dimensionality and enhanced fluctuations reveal novel magnetic phenomena~\cite{ghosh2021low,liang2023small,wu2023delaminated,ghosh2024electronic}.
Among these, 2D antiferromagnets (AFMs) are of particular interest due to their unique properties, such as ultrafast dynamics, minimal stray fields, and robustness against external magnetic fields, which makes them attractive for information storage and processing~\cite{baltz2018antiferromagnetic,gomonay2014spintronics}. 
In particular, transition-metal phosphotrichalcogenides, denoted MPX\textsubscript{3} (usually with M = Fe, Ni, Mn and X = S, Se) are van der Waals-layered AFM semiconductors with interesting magnetic and spintronic properties~\cite{mayorga2017layered,wang2018new,samal2021two,ghosh2021low}.
Understanding the spin dynamics and the role of thermal fluctuations in these low-dimensional systems is crucial, as their low dimensionality enhances thermal fluctuation. At the same time, their intrinsic uniaxial anisotropy stabilizes the AFM order. 

The dynamics of magnetization in magnetic systems is typically described by the Landau-Lifshitz-Gilbert (LLG) equation, which accounts for precession and damping of the magnetization under the action of an external magnetic field~\cite{gilbert1955lagrangian,gilbert2004phenomenological,kamra_prb_2018,brown_prev_1963}. 
In AFM materials, the structure made of two sublattices leads to coupled LLG equations that in turn give rise to interesting physical effects, \eg mode-dependent damping and spin-pumping~\cite{liu2017mode, kamra_prb_2018}.
At finite temperature, one can extend the LLG equation to account for thermal fluctuations by incorporating Langevin random fields into the LLG equation, producing a stochastic LLG equation~\cite{garanin1997fokker,chotorlishvili_prb_2013,atxitia_prb_2012}.

In this work, starting from a stochastic LLG equation, we develop and apply a comprehensive Fokker-Planck (FP) formalism tailored for describing fluctuations and dynamics in an AFM system with uniaxial anisotropy. Our primary goal is to use the FP equation, in a mean-field (MF) approximation, to derive the equations of motion for the spin polarization and the two-time spin-spin correlation functions. This approach allows us to incorporate thermal fluctuation corrections.

The results of this formalism are exploited to analyze the spin-wave dynamics, providing a renormalization of characteristic energies and damping rates induced by thermal fluctuations in the AFM phase.
Moreover, we have developed a phenomenological model to describe the influence of magnetic fluctuations on the resistance noise of 2D antiferromagnetic semiconductors,
motivated by recently observed anomalous resistance fluctuations in MPX$_3$ near the N\'eel temperature~\cite{ghosh2021low,ghosh2024electronic}.

This paper is structured as follows: Section~\ref{sec:model} presents a classical model for an AFM system. Here, starting from the stochastic LLG equation, we derive the FP equation to describe the evolution of the spin probability distribution function. 
These results are used in Section~\ref{sec:EOM}, where we solve for the stationary spin distribution and then for the two-time spin-spin correlation functions. 
In Section~\ref{sec:applications}, we discuss the application to spin-wave dynamics, and we detail the phenomenological model for resistance fluctuations.  Finally, conclusions are drawn in Section~\ref{sec:conclusions}.

\section{Model}
    \label{sec:model}

In this Section, we first present a classical model for describing a simple AFM with uniaxial anisotropy~\cite{rezende2019introduction}. We then introduce a stochastic Landau-Lifshitz equation derived from this model, in which Langevin fields are incorporated to account for thermal fluctuations. This stochastic approach leads to a FP formalism that describes the staggered magnetization dynamics under the influence of thermal fluctuations.

    \subsection{Magnetism}
    \label{subsec:magnetization}
    \begin{figure*}
    \begin{overpic}[width=0.8\textwidth]{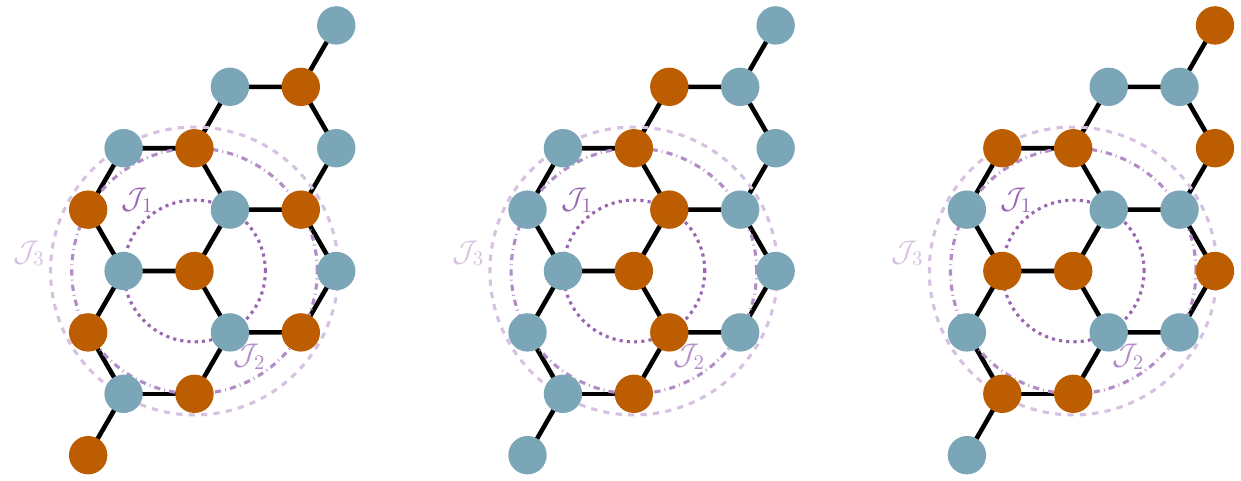}\put(6,32){ a)}\put(41.5,32){ b)}\put(77,32){ c)}\end{overpic}
        \caption{ Schematics of the three antiferromagnetic ordering patterns in a single-layer honeycomb lattice: a) N\'eel, b) zigzag, and c) stripy configurations. Orange and blue circles denote opposite spin orientations. The purple lines indicate the nearest (dotted line), second-nearest (dash-dotted line), and the third-nearest-neighbor  sites (dashed line) of the central site.}
        \label{fig:honeycomb}
    \end{figure*}

We model the magnetic properties of a two-dimensional antiferromagnet by an effective classical spin Hamiltonian defined on a lattice~\cite{chittari2016electronic},
\begin{equation}
\begin{aligned}
\label{eq:general_hamiltonian}
{\cal H} =& -\sum_{j} \bm{h}_{j}^{\mathrm{ext}} \cdot \bm{s}_{j}
+ \sum_{\langle j,k\rangle} \left( \mathcal{J}_1 \bm{s}_{j} \cdot \bm{s}_{k}
+ \mathcal{D}_1 s_{j,z}s_{k,z} \right) \\
&+ \sum_{\langle\langle j,k\rangle\rangle} \left( \mathcal{J}_2\bm{s}_{j} \cdot \bm{s}_{k}
+ \mathcal{D}_2 s_{j,z}s_{k,z} \right) \\
&+ \sum_{\langle\langle\langle j,k\rangle\rangle\rangle}
\left( \mathcal{J}_3\bm{s}_{j} \cdot \bm{s}_{k}
+ \mathcal{D}_3 s_{j,z}s_{k,z} \right)~.
\end{aligned}
\end{equation}
Here, the unit vector $\vb*s_{j}$ specifies the direction of the spin at lattice site $j$. The first contribution is the Zeeman term, which describes the coupling to the external magnetic field $\bm{h}^{\rm ext}$, while the remaining contributions describe the exchange interaction energy, which we take to be uniform.
In particular,  $\mathcal{J}_1$, $\mathcal{J}_2$, and $\mathcal{J}_3$ are the exchange coupling constants
between nearest (NN), second-nearest (SN), and third-nearest-neighbor (TN) spins, denoted as $\langle \cdot,\cdot \rangle$, $\langle \langle\cdot,\cdot \rangle\rangle\rangle$, $\langle \langle\langle\cdot,\cdot \rangle\rangle\rangle$  respectively.
The associated parameters $\mathcal{D}_1$, $\mathcal{D}_2$, and $\mathcal{D}_3$ represent the axial coupling constants, taking into account the fact that the system can have a cylindrical symmetry due to the geometry of the configuration, so they describe the presence of a strong easy-axis or easy-plane anisotropy.
This type of uniaxial term explicitly breaks spin-rotational symmetry and can play a crucial role in stabilizing long-range 2D magnetic order~\cite{rezende2019introduction,olsen_2dmaterials_2024}.
The effective field acting on a given spin at the lattice site $j$
        \begin{equation}\label{eq:eff_general}
        \begin{aligned}
        \bm{h}_{j}=-\frac{\partial \cal H}{\partial \bm{s}_{j}}&=\bm{h}^{\rm ext}_{j}-[({\cal J}_1\sum\nolimits'_k +{\cal J}_2\sum\nolimits''_k+{\cal J}_3\sum\nolimits'''_k ) \bm{s}_{k}]\\
        &-[({\cal D}_1\sum\nolimits'_k +{\cal D}_2\sum\nolimits''_k+{\cal D}_3\sum\nolimits'''_k ) s_{k,z}] \bm{u}_z~,
        \end{aligned}
        \end{equation}
where $\bm{u}_z$ denotes the unit vector directed along the $z$-direction, and $\sum\nolimits'_k$, $\sum\nolimits''_k$, and $\sum\nolimits'''_k$ represent, respectively, the sums over the NN, SN, and TN sites $k$ associated with site $j$.
The Hamiltonian given in Eq.~\eqref{eq:general_hamiltonian} has a quite general form, enabling the description of AFM systems with various spin-compensated ordering patterns. Fig.~\ref{fig:honeycomb} illustrates three  
possible spin-ordering arrangements for an antiferromagnetic system on a 2D honeycomb lattice, representative of the MPX$_3$ materials~\cite{chittari2016electronic}. The energetically favored configuration depends on the specific values of the interaction parameters ${\cal J}_i$ and ${\cal D}_i$ with $i = 1,2,3$.
In each case, we can decompose the full lattice into two sublattices, denoted $A$ and $B$. Accordingly, we replace the lattice index $j$, which can take $2N$ values, with a pair $(\alpha, p)$, where $\alpha \in \{A, B\}$ and $p$ runs over $N$ values.
If the exchange interaction dominates over the anisotropies and external fields, the spatial variations are smooth, and the thermal fluctuations are small, we can apply the macrospin approximation~\cite{rezende2019introduction}. 
For each sublattice, we have a collective magnetization, $\vb* M_{\alpha }= N_\alpha \vb*s_{\alpha}$, where $ N_\alpha$ being the spin density on the sublattice $\alpha \in \{A,B\}$, and $N_A=N_B=N$.
Neglecting the external magnetic field and adopting the macrospin approximation, the effective field acting on the spins of sublattice $\alpha$ becomes spatially uniform and can be expressed as
\begin{equation}
\label{eq:eff-h-general}
\bm{h}_{\alpha }=-J \bm{s}_{\bar{\alpha}}-\Delta s_{\bar\alpha,z} \bm{u}_z
-J' \bm{s}_{\alpha}-\Delta' s_{\alpha,z} \bm{u}_z
~,
\end{equation}
where $\bar{\alpha}=B$ ($\bar{\alpha}=A$) when $\alpha=A$ ($\alpha=B$). The coefficients $J$ and $J'$ ($\Delta$ and $\Delta'$) are given as linear combinations of the interaction parameters ${\cal J}_i$ (${\cal D}_i$), specifically
        \begin{subequations}
        \begin{align}
        J&=(Z_1-Y_1) {\cal J}_1+(Z_2-Y_2) {\cal J}_2+(Z_3-Y_3) {\cal J}_3~,\\
        J'&=Y_1 {\cal J}_1+Y_2 {\cal J}_2+Y_3 {\cal J}_3~,\\     
        \Delta&=(Z_1-Y_1) {\cal D}_1+(Z_2-Y_2) {\cal D}_2+(Z_3-Y_3) {\cal D}_3~,\\
        \Delta'&=Y_1 {\cal D}_1+Y_2 {\cal D}_2+Y_3 {\cal D}_3~,  
        \end{align}
        \end{subequations}
with $Z_1$, $Z_2$, and $Z_3$ denoting, respectively, the number 
of NN, SN, and TN sites with respect to an arbitrary one (independently of the sublattice), and $Y_1$, $Y_2$, and $Y_3$ representing the corresponding number
of NN, SN, and TN sites belonging to the same sublattice.
As an illustrative example, we consider the honeycomb lattice shown in Fig~\ref{fig:honeycomb}, where one has $(Z_1,Z_2,Z_3)=(3,6,3)$. Here, the set $(Y_1,Y_2,Y_3)$ depends on the ordering pattern, taking the values $(0,Z_2,0)$, $(2,2,0)$, and $(1,2,3)$ for the configurations N\'eel, zig-zag, and stripy, respectively.

Within this classical description, the energetically preferred AFM configuration is obtained by maximizing the magnitude of $\bm{h}_{\alpha}$ under the constraint that it points in the direction opposite to $\bm{s}_{\alpha}$ and parallel to $\bm{s}_{\bar \alpha}$.
For simplicity, we henceforth restrict our analysis to nearest-neighbor exchange interactions, ${\cal J}_1$ and ${\cal D}_1$, noting that the formalism developed below can be straightforwardly generalized to include further-neighbor couplings. 
In particular, we focus on the parameter regime $ {\cal J}_1>0$ and ${\cal J}_1+{\cal D}_1>0$, which stabilize an AFM configuration. For a single honeycomb layer, taken as a representative example, this condition clearly stabilizes a N\'eel configuration.
 In particular, for $\mathcal D_1 > 0$, this model favors alignment along the $z$ direction. Conversely, for $\mathcal D_1  <  0$ one obtains a stronger interaction in the $x$-$y$ plane. The spherical symmetry is recovered by taking $\mathcal D_1=0$. 
For the sake of clarity, and within the restriction stated above, we explicitly write the Hamiltonian for the two interpenetrating classical spin sublattices ($A,B$)~\cite{chen2019landau,rezende2019introduction},
        \begin{equation}
        \begin{aligned}
        \label{eq:starting hamiltonian}
            \mathcal H &=-\sum_{\alpha= A,B} \sum_p\bm{h}^{\rm ext}_{{\alpha p}} \cdot  \vb*s_{\alpha p}+{\mathcal J}_1 \sum_{\expval{p,q}} \vb*s_{A p} \cdot\vb*s_{B q} \\
            &+ {\mathcal D}_1 \sum_{\expval{p,q}} s_{A p,z} s_{B q,z}~,
        \end{aligned}
        \end{equation}
and the effective field acting on a given spin at site $p$ of sublattice $\alpha$ is given by
        \begin{equation}\label{eq:eff}
        \bm{h}_{\alpha p}=-\frac{\partial \cal H}{\partial \bm{s}_{\alpha p}}=\bm{h}^{\rm ext}_{\alpha p}-{\cal J}_1\sum\nolimits'_q \bm{s}_{\bar{\alpha} q}-{\cal D}_1 \sum\nolimits'_q  s_{\bar{\alpha}q, z} \bm{u}_z~.
        \end{equation}
        When disregarding the external magnetic field and applying the macrospin approximation, the effective field assumes a homogeneous form and can be rewritten as
        \begin{equation}
        \label{eq:eff-h}
        \bm{h}_{\alpha }=-J \bm{s}_{\bar{\alpha}}-\Delta s_{\bar\alpha,z} \bm{u}_z~,
        \end{equation}
        where we recall $J=Z_1 {\cal J}_1$ and $\Delta=Z_1 {\cal D}_1$.
Before concluding, we briefly outline how corrections due to long-wavelength spatial inhomogeneities can be added to the macroscopic approximation.
Starting from Eq.~\eqref{eq:eff}, in the absence of the external field, and assuming a generic lattice, we make use of its discrete translational invariance to rewrite the effective field in terms of its Fourier transform as
\begin{equation}
\begin{aligned}
\tilde{\bm{h}}_{\alpha}(\bm{P})&=\frac{1}{N}\sum_{i} e^{-i\bm{P} \cdot \bm{R}_p} \bm{h}_{\alpha p}\\
&=-f(\bm{P})[{\cal J}_1  \tilde{\bm{s}}_{\alpha}(\bm{P})+{\cal D}_1  \tilde{s}_{\alpha z}(\bm{P})\bm{u}_z]
\end{aligned}
\end{equation}
where $f(\bm{P})=\sum_{l=1}^{Z_1} e^{-i\bm{P} \cdot \bm{d}_l}$ denotes the form factor, and $\{\bm{d}_l\}_{l=1}^N$ represent the NN vectors.
As an example, we focus on the honeycomb lattice, for which $\bm{d}_l = a_0(\sin(l\pi/3),\cos(l\pi/3))^{\rm T}$~\cite{pellegrino_prb_2009}. We work in the long-wavelength limit, valid when the characteristic length scale of the inhomogeneities is much larger than the lattice spacing $a_0$. In this regime, the form factor can be expanded for small $P$, $P \ll 1/a_0$, and approximated as
$
f(\bm{P}) \approx 3(1-a_0^2 P^2/4)
$.

Under this long-wavelength approximation, we now transform back to real space, so that
        \begin{equation}
        \label{eq:eff-h-realspace}
        \bm{h}_{\alpha }(\bm{r})=-J\bigg(1+\frac{a_0^2}{4}\nabla^2\bigg) \bm{s}_{\bar{\alpha}}(\bm{r})-\Delta \bigg(1+\frac{a_0^2}{4}\nabla^2\bigg) s_{\bar\alpha,z}(\bm{r}) \bm{u}_z~,
        \end{equation}
where we have  the replacement $\bm{P}\to i\nabla$, with $\nabla$ representing the nabla operator.
It is straightforward to verify that the above expression remains valid for a 2D square lattice, with the form factor $f(\bm{P}) = 2[\cos(P_x a_0) + \cos(P_y a_0)]$. For a three-dimensional cubic lattice, however, the prefactor is reduced, with $a_0^2/4$ being replaced by $a_0^2/6$~\cite{garanin1997fokker}.
From here on, for the sake of simplicity, we employ the FP formalism within the macrospin approximation. Nevertheless, the discussion presented here could be useful for future extensions that explicitly account for spatial inhomogeneities.

    \subsection{Fokker-Planck formalism}
        \label{subsec:FP equation}

        Here, we derive the dynamics equation for the probability distribution of the magnetization configuration of an AFM, starting
        from a stochastic version of the LLG equation that includes Langevin fields to incorporate the effects of thermal fluctuations ~\cite{garanin1997fokker}.
        For each of the two sublattices constituting the AFM, the stochastic LLG equations are expressed as
        \begin{equation}
            \label{eq:LLG_AFM}
            \dot{\vb*s}_{\alpha p} = \vb*s_{\alpha p} \times \qty(\vb*h_{\alpha p}+\bm{\zeta}_{\alpha p} -  \Lambda  \dot{\vb* s}_{\alpha p}  -  \Lambda' \dot{\vb* s}_{\bar \alpha p} )~,
        \end{equation}
        where  $\dot{\vb* s}_{\alpha p} =\pdv {\vb*s_{\alpha p}}{t} $,  $\alpha\in\{A,B\}$, $\bm{h}_{\alpha p}$ is the effective magnetic field associated with the Hamiltonian of the spin defined in Eq.~\eqref{eq:eff}, $\bm{\zeta}_{\alpha p}$ is a Langevin field that is postulated to be uncorrelated at different sites. Note that we are working in units such that the lattice-independent~\cite{chen2019landau}, gyromagnetic ratio $\mu=1$, and $\hbar=1$.
        Both $\Lambda$ and $\Lambda'$ are two diagonal matrices
        \begin{subequations}
        \begin{align}
        \Lambda&=\diag(\lambda+\delta \lambda,\lambda+\delta\lambda,\lambda)~,\\
        \Lambda'&=\diag(\lambda+\delta\lambda,\lambda+\delta\lambda,\lambda)~,      
        \end{align}
        \end{subequations}
         where $\lambda,\lambda+\delta \lambda \ll 1$ are the intra-sublattice Gilbert damping factors~\cite{gilbert1955lagrangian,gilbert2004phenomenological}, 
         and $\lambda',\lambda'+\delta \lambda'\ll 1$ are the inter-sublattice Gilbert damping factors~\cite{kamra_prb_2018,kamra_prl_2017}.
         Inclusion of parameters $\delta \lambda$ and $\delta \lambda'$
         allows uniaxial symmetry in damping mechanisms, following the cylindrical symmetry of the spin-spin interactions. 

       By linearizing this pair of equations in the Gilbert damping factors, and considering the Langevin fields weak with respect to the effective field, $|\vb*\zeta_{\alpha p}|\ll |\vb*h_{\alpha p}| $, the expression in Eq.~\eqref{eq:LLG_AFM} turns into
        \begin{equation}\label{eq:LLG-Langevin}
            \dot{\vb* s}_\alpha = \vb* s_\alpha \times (\vb* h_\alpha + \vb* \zeta_\alpha)  
                - \Lambda  \vb* s_\alpha \times (\vb* s_\alpha \times \vb* h_\alpha) 
                - \Lambda'  \vb* s_\alpha \times (\vb* s_{\bar{\alpha}} \times \vb* h_{\bar{\alpha}}) ~.
        \end{equation}
        Within the macrospin approximation, spatial fluctuations of the site degrees of freedom are neglected.
        From here on out, we omit the site index.
        
        Following Ref.~\cite{garcia2007introduction}, Eq.~\eqref{eq:LLG-Langevin} can be rewritten in the form of the (componentwise) Langevin equation
        \begin{equation}\label{eq:motion}
          \dot {s}_{\alpha i}= \mathcal A_{\alpha i} + \mathcal B_{\alpha i, \beta k}  \zeta_{\beta k}~,
        \end{equation}
        where the drift term $\mathcal{ A}_{\alpha i}$ and the diffusion term $\mathcal B_{\alpha i, \beta k}$ are identified as
        \begin{subequations} \label{eq:coeff for KME}
        \begin{align}
                \mathcal{ A}_{\alpha i}&= 
                \epsilon_{i j k} s_{\alpha j} h_{\alpha k} 
                - \Lambda_{ii} [s_{\alpha j} h_{\alpha j} s_{\alpha i}-h_{\alpha i}]\\
                & - \Lambda_{ii}' [s_{\alpha j} h_{\bar{\alpha} j} s_{\bar{\alpha} i}-s_{\alpha j}  s_{\bar{\alpha} j} h_{\bar{\alpha} i}] \nonumber~,\\
                \mathcal B_{\alpha i, \beta k} &=  \delta_{\alpha \beta} \epsilon_{i j k} s_{\alpha j}~,
        \end{align}        
        \end{subequations}
        where we have used the fact that $\bm{s}_\alpha$ is a unit vector.
        Note that we used Einstein's summation rule on the repeated indices and that we used Greek letters to label the sublattices and Latin letters when referring to the Cartesian coordinates $\{x,y,z\}$, and $\delta_{\alpha \beta}$ is the Kronecker's delta symbol. 

        Langevin fields $\bm{\zeta}_\alpha$ are considered stochastic processes~\cite{titov_jmm_2021,garcia2007introduction,garanin1997fokker}. Assuming that they are Gaussian and that they satisfy the following statistical properties
        \begin{subequations}\label{eq:zeta}
        \begin{align}
        \expval{\vb* \zeta_{\alpha}(t)}&=\bm{0}~,\\
        \expval{\zeta_{\alpha j}(t_1)\zeta_{\alpha k}(t_2)}&=2D_{\mathrm T,j} \delta_{\alpha \beta}\delta_{jk}\delta(t_1 - t_2)~. \label{eq:DTj}
        \end{align}
        \end{subequations}
        Here, we assume that the diffusion coefficients have cylindrical symmetry following the anisotropy of the Hamiltonian, and are identical for both sublattices, namely 
        $\vb* D_\mathrm{T}=(D_\mathrm{T},D_\mathrm{T},D_\mathrm{T} +\delta D_\mathrm{T})^\mathrm{T} $.

        Moreover, we underline that the condition $D_{\mathrm{T},j}  \geq  0$ leads to a constraint on the anisotropy $\delta D_\mathrm{T}  \geq  -D_\mathrm{T}$.
        Using the quantities in Eqs.~\eqref{eq:coeff for KME} and~\eqref{eq:zeta}, it is possible to obtain the coefficients of the Kramers-Moyal expansion, which describes the evolution of the probability distribution function (PDF) for a generic Markov process~\cite{kramers1940brownian,moyal1949quantum,garcia2007introduction} (see Appendix~\ref{app:lambda'} for further details). 
        One obtains the FP equation~\cite{garcia2007introduction}, namely the equation governing the dynamics for the PDF, $\mathcal P (\vb*{\mathcal S},t)$,
        \begin{widetext}
        \begin{equation}
        \label{eq:FP equation}
             \dot{\mathcal P} (\vb*{\mathcal S},t)=- \pdv{\vb* s_\alpha} \cdot \big\{ \vb* s_\alpha \times {\vb* h}_{\alpha} 
             + D_{\mathrm T} {\vb* s}_\alpha \times (\vb* s_\alpha \times \partial_{\vb* s_\alpha})
             + \delta D_{\mathrm T} {\vb* s}_\alpha \times [(\vb* s_\alpha \times \partial_{\vb* s_\alpha})_z \vb*u_z]
             -\Lambda \; \vb* s_\alpha \times (\vb* s_\alpha \times {\vb* h}_{\alpha}) 
             -\Lambda' \; \vb* s_\alpha \times (\vb* s_{\bar \alpha} \times {\vb* h}_{\bar \alpha}) 
             \big\} \mathcal P (\vb*{\mathcal S},t)~,
        \end{equation}
        \end{widetext}
        that represents the probability of having the spin configuration $\vb*{\mathcal S}=(\vb*s_\mathrm{A}, \vb*s_\mathrm{B})$ at time $t$, and we used Einstein's summation rule over the sublattice index $\alpha$.

\section{Equations of motion}
    \label{sec:EOM}

     In this Section, starting from the FP equation shown in Eq.~\eqref{eq:FP equation}, we derive the equation of motion for the spin polarization, identified as the first moment~\cite{garanin1997fokker}, and for the two-time spin-spin correlation function~\cite{likharev2019statistical,kogan2008electronic}.

    \subsection{Spin polarization}
    \label{sec:stationary}
        The spin polarization for a specific sublattice $\alpha \in \{A,B\}$ is described in terms of the PDF as 
        \begin{equation}\label{eq:malpha}
        \vb* m_\alpha = \expval{\vb*s_\alpha} = \int \dd[6]\vb*{\mathcal S} \, \mathcal P(\vb*{\mathcal S}, t) \vb*s_\alpha~,
        \end{equation}
        where we introduced the notation $\int \dd[6] \vb*{\mathcal S} = \int \dd[3]\vb*s_{A} \int \dd[3]\vb*s_{B}$.
        To derive the equation of motion of the variable $\vb* m_\alpha$,  we consider the time derivative on both sides of the aforementioned definition.
        Using the result of Eq.~\eqref{eq:FP equation} and introducing the following notation $\expval{\star} = \int \dd[6]\vb*{\mathcal S} \, \mathcal P(\vb*{\mathcal S}, t) \star$, we find that the equation governing the dynamics of the spin polarization can be written in the following form:
        \begin{equation}
        \label{eq:dynamics magnetization vector}
        \begin{aligned}
            \dot{\vb*m}_\alpha =& \expval{\vb* s_\alpha \times \vb* h_\alpha} \\
                &- 2 D_{\mathrm T} \vb* m_{\alpha} - 2  \delta D_{\mathrm T} (m_{\alpha x} \vb*u_x +m_{\alpha y} \vb*u_y) \\
                &- \Lambda \expval{\vb* s_\alpha \times(\vb* s_\alpha \times \vb* h_\alpha)}~,
        \end{aligned}
        \end{equation}
        where on the right-hand side the first line represents the precession term, the second the diffusion term, and the third the Gilbert damping term.
        For the sake of clarity, we exclude the inter-sublattice Gilbert term $\Lambda'$ from the present discussion. The full derivation of the equation of motion of the spin polarization, including $\Lambda'$, is provided in Appendix~\ref{app:lambda'}.
        %
        Note that the anisotropy terms $\delta D_{\rm T}$ and $\delta \lambda$ only affect the in-plane component of the magnetization, and that the evaluation of the time derivative of the first-order moment involves higher-order moments (namely, second and third).
        The difficulty of solving the FP equation is transferred to the generation of an infinite hierarchical set of equations, which involves the higher-order moments.
        To solve this, in analogy to the approach used in~\cite{garanin1997fokker} for a ferromagnetic classical model, we apply the MF approximation. 
       The PDF $\mathcal P_{\mathrm{MF}}$ assumes that the spins of the sublattices, $\vb*s_A$ and $\vb*s_B$, are independent. 
       Each one is coupled to an effective external field $k_{\rm B} T\vb* \xi_\mathrm{A}$ and $k_{\rm B} T\vb* \xi_\mathrm{B}$, respectively, where $k_{\rm B} T$ is the thermal energy~ \cite{garanin1997fokker}. The PDF $\mathcal P_{\mathrm{MF}}$ is expressed as
        \begin{equation}\label{eq:PMF}
            \mathcal P_{\mathrm{MF}} = \frac {e^{\vb* \xi_\mathrm{A} \cdot \vb*s_\mathrm{A}}}{\mathcal Z(\xi_A)}
                            \frac {e^{\vb* \xi_\mathrm{B} \cdot \vb*s_\mathrm{B}}}{\mathcal Z(\xi_B)} ,
        \end{equation}
        where $\mathcal Z(\xi_\alpha)=\int \dd[3]\vb*s_{\alpha} e^{\vb* \xi_\alpha \cdot \vb*s_\alpha} =4\pi \sinh(\xi_\alpha)/\xi_\alpha$ is the partition function.
        As a consequence, the spin polarization is expressed as
        \begin{equation}\label{eq:MFm}
        {\bm m}_\alpha=\frac{B(\xi_\alpha)}{\xi_\alpha} {\bm \xi}_\alpha~,
        \end{equation}
        where $B(\xi)$ denotes the Langevin function~\cite{langevin1905}
        \begin{equation}\label{eq:langevin}
        B(x)\equiv \coth(x)-\frac{1}{x}~.
        \end{equation}
        Similarly, calculating any term of Eq.~\eqref{eq:dynamics magnetization vector}, one finds the following equation
        \begin{equation}
        \label{eq:LLB}
        \begin{aligned}
        \dot{\bm{m}}_\alpha&=\bm{m}_\alpha \times \expval{{\bm h}_\alpha}_{\rm MF}\\
        &- 2 D_{\mathrm T} \vb* m_{\alpha} - 2  \delta D_{\mathrm T} (m_{\alpha x} \vb*u_x +m_{\alpha y} \vb*u_y) \nonumber\\
        &- \Lambda {h}_{\alpha j}{\expval{s_{\alpha j} \bm{s}_\alpha  }}_{\rm MF}
        +\Lambda  \expval{{\bm h}_\alpha}_{\rm MF}~,
        \nonumber 
        \end{aligned}
        \end{equation}
        where we used Einstein’s summation rule on repeated Cartesian coordinates, $j \in \{x,y,z\}$, 
        \begin{subequations}
        \begin{align}
        \expval{s_{\alpha i} s_{\alpha j}}_{\rm MF}&= \frac{\xi_{\alpha i}\xi_{\alpha j}}{\xi_{\alpha}^2} + \frac{B(\xi_\alpha)}{\xi_\alpha}\left(\delta_{ij}-3\frac{\xi_{\alpha i}\xi_{\alpha j}}{\xi_{\alpha}^2}\right)~, \label{eq:ss}\\
        \expval{s_{\alpha i} s_{\bar{\alpha} j}}_{\rm MF}&=  \expval{s_{\alpha i}}_{\rm MF}\expval{ s_{\bar{\alpha} j}}_{\rm MF}~, \label{eq:sbs}\\
        \expval{{\bm h}_\alpha}_{\rm MF}&=-J \bm{m}_{\bar{\alpha}}-\Delta m_{\bar{\alpha} z} \bm{u}_z~,
        \end{align}
        \end{subequations}
        and $\bar{\alpha}=B$ ($\bar{\alpha}=A$) for $\alpha=A$ ($\alpha=B$).

        Now, we focus on the stationary limit $\dot{\bm m}_\alpha=0$, and impose the AFM order $\bm{m}_A=-\bm{m}_B$ and $\Delta>0$.
        Assuming that the spins are aligned along the $z$ direction, such that $\vb*\xi_\mathrm{A}=-\vb*\xi_\mathrm{B}=\xi \vb*u_z$, where $\xi$ solves
        \begin{equation}\label{eq:static_B}
        \frac{B(\xi)}{\xi}=\frac{D_{\rm T}}{ \lambda (J+\Delta)}~. 
        \end{equation}
       By imposing 
       \begin{equation}\label{eq:DTthermal}
       D_{\rm T}=\lambda k_{\rm B} T~,
       \end{equation}
       Eq.~\eqref{eq:static_B} coincides with the nonlinear equation that determines the global minimum of the free energy via the classical Hamiltonian, as expressed in Eq.~\eqref{eq:starting hamiltonian}, within the MF approximation~\cite{goldenfeld_book}.
        %
       Proceeding with the assumption that there is a stationary solution with spins lying in the $x$-$y$ plane, \eg along the $x$ direction such that $\vb*\xi_\mathrm{A}=-\vb*\xi_\mathrm{B}=\xi \vb*u_x$, the resulting expression for $\xi$ is given by       
        \begin{equation}\label{eq:static_Bx}
        \frac{B(\xi)}{\xi}=\frac{D_{\rm T}+\delta D_{\rm T}}{ (\lambda+\delta \lambda) J}~. 
        \end{equation}
        Similarly to the above analysis, this is equivalent to the nonlinear equation that identifies a local minimum of the free energy via the classical Hamiltonian in the MF approximation $B(\xi)/\xi=k_{\rm B} T/J$, assuming $ D_{\rm T} +\delta D_{\rm T} = (\lambda+\delta \lambda) k_{\rm B} T$. 
        Combining this result with Eq.~\eqref{eq:DTthermal}, one has
        \begin{equation}\label{eq:dDTthermal}
        \delta D_{\rm T}=\delta \lambda ~k_{\rm B}T~.
        \end{equation}

        Henceforth, we assume that the spins are oriented along the $z$ axis, and we will enforce Eqs.~\eqref{eq:static_B} and \eqref{eq:DTthermal}, thus establishing the following identity
        \begin{equation}\label{eq:self}
                \frac{B(\xi)}{\xi}=\frac{k_{\rm B} T}{J+\Delta}~, 
        \end{equation}
        and $\Delta>0$. 
        %
        %
        Furthermore, $B(\xi)/\xi$ is a monotonic function of $\xi$, and it is bounded, \ie $0<B(\xi)/\xi\le 1/3$.
        So, Eq.~\eqref{eq:self}  has a solution for only temperatures $T\le T_{\rm N}$, where
        \begin{equation}\label{eq:TN}
        T_{\rm N}=\frac{J+\Delta}{3 k_{\rm B}}~
        \end{equation}
        is known as N\'eel temperature and implicitly defines $\xi(T)$ as a function of $k_{\rm B} T/(J+\Delta)$.
        We conclude with the analysis of the asymptotic behaviors of $\xi(T)$: as $T\to 0$, $\xi$ reaches large values, behaving as $\xi \approx 3 T_{\rm N}/T$. On the other hand, as $T$ approaches $T_{\rm N}^-$, $\xi$ assumes  small values that are approximated by $\xi \approx \sqrt{15} \sqrt{1-T/T_{\rm N}}$.
        
        
    \subsection{Spin-spin correlation functions}
    \label{sec:2-time correlators}

        Here, we aim to find an expression for the correlation between 
        the fluctuations of the spin vector on sublattice $\beta$ at time $t_0$ around its stationary value, $\vb*s_\beta (t_0) - \vb*m_\beta$, and the fluctuations of the spin vector on sublattice $\alpha$ at a later time $t_1>t_0$, $\vb*s_\alpha (t_1) - \vb* m_\alpha$, in the stationary regime, namely:
        \begin{widetext}
        \begin{equation}
        \label{eq:2-time correlator base}
            g_{\alpha i, \beta \ell}(t_1,t_0)=\expval{[s_{\alpha i}(t_1)- m_{\alpha i}] [s_{\beta \ell}(t_0) - m_{\beta \ell}]} =
            \int \dd[6]{\vb*{\mathcal S}} \int \dd[6]{\vb*{\mathcal S}'}\, [s_{\alpha i}- m_{\alpha i}] [s_{\beta \ell}' - m_{\beta \ell}]\,
            \mathcal P (\vb*{\mathcal S}, t_1 | \vb*{\mathcal S}', t_0) \mathcal P_0(\vb*{\mathcal S}')
            ~,
        \end{equation}
        \end{widetext}
        where the indices $i$ and $\ell$ refer to the Cartesian coordinates $x,y,z$, $\mathcal P (\vb*{\mathcal S}, t_1 | \vb*{\mathcal S}', t_0)$ is the conditional probability of having the spin configuration $\vb*{\mathcal S}=(\vb*s_\mathrm{A}, \vb*s_\mathrm{B})$ at time $t_1$ assuming that the configuration at time $t_0$ was $\vb*{\mathcal S}'=(\vb*s'_\mathrm{A}, \vb*s'_\mathrm{B})$, $P_0(\vb*{\mathcal S}')$ is the stationary probability distribution~\cite{likharev2019statistical}.
        Since we enforce the AFM condition $\bm{m}_A = -\bm{m}_B$ with $\Delta > 0$, and we set $\vb*\xi_\mathrm{A}=-\vb*\xi_\mathrm{B}=\xi \vb*u_z$.
        In the stationary regime, the correlation functions depend only on the time difference, so that $g_{\alpha i, \beta \ell}(t_1,t_0) \to g_{\alpha i, \beta \ell}(t)$, with $t = t_1 - t_0$ for simplicity.

       By differentiating Eq.~\eqref{eq:2-time correlator base} with respect to time through the use of Eq.~\eqref{eq:FP equation}, we derive the equation governing the dynamics of the correlation functions. The complete derivation is detailed in Appendix~\ref{app:two-time correlators}, where it is shown that achieving an exact resolution of the motion equation requires solving a hierarchical series of equations. 
       Since this hierarchy is infinite, an exact solution is unfeasible. Therefore, to derive a closed expression for the correlation functions, it becomes necessary to truncate the hierarchy. In the spirit of the Gaussian approximation, the truncation is performed by assuming that Wick's theorem is applicable to the correlation functions of fluctuations of order higher than second~\cite{goldenfeld_book}, allowing them to be expressed in terms of second-order correlation functions.
       Initial conditions are fixed considering that in the limit $t \to 0$,
       we have $\mathcal P (\vb*{\mathcal S}, t | \vb*{\mathcal S}', 0) \rightarrow \delta(\vb*{\mathcal S}-\vb*{\mathcal S}')$. It follows that
        \begin{equation}
        \label{eq:initial_condition}
            g_{\alpha i, \beta \ell}(0) = \expval{s_{\alpha i}(0)s_{\beta \ell}(0)}_{0} - m_{\alpha i}m_{\beta \ell} = \delta_{\alpha, \beta } \delta_{i,\ell} \sigma_i^2~,
        \end{equation}
        where
        \begin{subequations}
         \begin{align}
                \sigma_{z} &= \sqrt{1-B^2(\xi) - 2 \frac{B(\xi)}{\xi}}~,\label{eq:sigmaz}\\    
        \sigma_{x}&=\sigma_{y}=\sigma_{\bot} =\sqrt{\frac{B(\xi)}{\xi}}~.\label{eq:sigmabot}
        \end{align}
        \end{subequations}
        %
        We underline that for temperatures $T\le T_{N}$, Eq.~\eqref{eq:static_B} holds, while for $T>T_{\rm N}$ 
        it is straightforward to see that $B(\xi)=0$ and $B(\xi)/\xi=1/3$.
        
        As a consequence of the symmetry of the system, the equations of motion for the correlation functions $ g_{\alpha i, \beta \ell}(t) $ naturally separate into in-plane components, associated with the Cartesian coordinates $x$ and $y$, and an out-of-plane component, involving only the $z$-components. In the following, we examine these two contributions independently, focusing on temperatures ranging from absolute zero to N\'eel temperature.

        \subsubsection{Out-of-plane dynamics}

            Following the derivation in Appendix~\ref{ssect:out},  the spin-spin correlation functions, which involve the out-of-plane dynamics, are expressed as
            \begin{equation}
            \label{eq:out-of-plane 2-time correlator}
            \begin{aligned} 
              g_{\alpha z, \beta z}(t) &= \Big[1-B^2(\xi)-2\frac{k_{\rm B} T}{J+\Delta}\Big] \Big[ \frac{e^{-\Gamma_\parallel |t|}}{2}\\
              &+(2\delta_{\alpha \beta}-1) \frac{e^{-\Gamma_\parallel' |t|}}{2} \Big]~,
             \end{aligned}        
            \end{equation}
            where the damping rates  are expressed as
            \begin{subequations}
            \label{eq:out-damping}
            \begin{align}
             \Gamma_\parallel&=2\lambda(J+\Delta)\Big[B^2(\xi)+\frac{2 k_{\rm B}T}{J+\Delta} \Big]~,   \\
            \Gamma_\parallel'&=2\lambda(J+\Delta)B^2(\xi)~.  
            \end{align}
            \end{subequations}
            Here, we have omitted the term $\lambda'$. In Appendix~\ref{app:two-time correlators}, it is shown that $\lambda'$ essentially results in a renormalization of the parameter $\lambda$, namely $\lambda\to\lambda\{1-\lambda'J/[\lambda(J+\Delta)]\}$.
            Note that an anisotropy in the diffusion term $\delta D_{\rm T}$ and the Gilbert damping terms $\delta \lambda$ and $\delta \lambda'$ do not affect the dynamics along the $z$ direction.

        \subsubsection{In-plane dynamics}

            According to the derivation shown in Appendix~\ref{ssect:in-plane}, the spin-spin correlation functions, which refer to the in-plane dynamics, are represented using the following block-matrix
            \begin{equation}
            \label{eq:in-plane dynamics 2-time correlators}
                \begin{pmatrix}
                    G_{xx}(t) & G_{xy}(t)  \\
                    G_{yx}(t)  & G_{yy}(t)  \\
                \end{pmatrix}
                =
                \frac{k_{\rm B} T}{J+\Delta}e^{-\Gamma_\bot |t|}
                \begin{pmatrix}
                    {\cal G}(|t|) & {\cal G}'(|t|)  \\
                    -{\cal G}'(|t|)  & {\cal G}(|t|)  \\
                \end{pmatrix}~,
            \end{equation}
           where
            \begin{subequations}
            \begin{align}
            \label{eq:in-plane dynamics 2-time correlators_blocks}
                G_{ij}(t) &=
                \begin{pmatrix}
                    g_{Ai,Aj}(t) & g_{Ai,Bj}(t)  \\
                    g_{Bi,Aj}(t)  & g_{Bi,Bj}(t)  \\
                \end{pmatrix}~,\\
            {\cal G}(t)  &=
                   \begin{pmatrix}
                    \cos(\Omega t)  & -\frac{\Gamma_\bot'}{\Omega}\sin(\Omega t)  \\
                    -\frac{\Gamma_\bot'}{\Omega}\sin(\Omega t)   & \cos(\Omega t)  \\
                \end{pmatrix}   ~,    \\
           {\cal G}'(t) &=
           B(\xi)
                   \begin{pmatrix}
                    \frac{J+\Delta}{\Omega}\sin(\Omega t)  &  \frac{J}{\Omega}\sin(\Omega t)   \\
                    - \frac{J}{\Omega}\sin(\Omega t)   & - \frac{J+\Delta}{\Omega}\sin(\Omega t)   \\
                \end{pmatrix}     ~.  
           \end{align}     
           \end{subequations} 
        Here, the damping factors have the following expressions
        \begin{subequations}
        \begin{align}
            \Gamma_\bot &=  (\lambda+\delta \lambda) (J+\Delta)  \Big[B^2(\xi) + (2+\phi_\bot) \frac{k_\mathrm{B} T}{J+\Delta} \Big]~,\label{eq:Gammabot}\\
            \Gamma_\bot' &=  (\lambda+\delta \lambda)  J \Big(1-\frac{k_\mathrm{B} T}{J+\Delta} \Big)~,\label{eq:Gammapbot}
        \end{align}
        \end{subequations}         
        where
        \begin{equation}\label{eq:phibot}
        \phi_\bot=-\frac{\delta \lambda}{\lambda+\delta \lambda}~,
        \end{equation}
        and the proper pulsation is
        \begin{equation}\label{eq:properOmega}
        \Omega=\sqrt{\Omega_0^2 B^2(\xi)-\Gamma_\bot'^2}~,
        \end{equation}    
which is expressed in terms of the zero-temperature spin-wave energy gap $\Omega_0=\sqrt{\Delta(2J+\Delta)}$~~\cite{rezende2019introduction}, and also incorporates a renormalization contribution stemming from the damping correction $\Gamma_\bot'$.
        %
        For the sake of clarity, the inter-sublattice terms $\lambda'$ and $\delta \lambda'$ have been omitted here. They have been examined in detail in Appendix~\ref{ssect:in-plane}, where it is shown how they are incorporated into $\Gamma_\bot$ and $\Gamma_\bot'$.

\section{Applications}
\label{sec:applications}
      Employing the method used to derive the equations of motion for spin polarization and spin-spin correlation functions, we examine two potential applications.
      Firstly, we investigate the dynamics of the spin-wave generated within the AFM phase~\cite{kamra_prb_2018}.
      Secondly, we show a phenomenological model tailored to describe the effects of magnetic fluctuations on the experimentally observed resistance fluctuations in MPX\textsubscript{3}~\cite{ghosh2024electronic}.

    \subsection{Spin-waves}
    \label{sec:spinwaves}
    
    Starting from the results of Sec.~\ref{sec:stationary}, we derive the equation of motion for spin polarization in the AFM configuration being perturbed by a weak, external  magnetic field $\bm{h}^{\rm ext}$ to both sublattices~\cite{gekht1976magnetic}.

    We start by analyzing the modifications in Eq.~\eqref{eq:dynamics magnetization vector} after substituting $\bm{h}_\alpha$ with $\bm{h}_\alpha+\bm{h}^{\rm ext}$, and using
    \begin{equation}
    {\cal P}={\cal P}_{0} [ 1 + \sum_{\alpha=A,B} \bm{w}_{\alpha} \cdot (\bm{s}_\alpha - \bm{m}_{\alpha}) ]~. \label{eq:Psw}
    \end{equation}
    In the stationary regime, where no external field is present, ${\cal P}_{0}$ represents the probability density function (PDF), and ${\bm m}_\alpha = {\expval{{\bm s}_\alpha}}_0$ denotes the spin polarization.
    Under the influence of a weak, time-dependent, external field, discarding terms $\order*{{h^{\rm ext}}^2}$, we find that  ${\bm w}_{\alpha} = \mathcal O (h^{\rm ext})$ depends on time. 
    Following the approach described in Sec.~\ref{sec:stationary}, we examine the stationary PDF achieved through the MF approximation, specifically ${\cal P}_{0} \rightarrow {\cal P}_{\rm MF}$.
    Here, we assume that ${\cal P}_{\rm MF}$ describes the stationary configuration in which the spins are oriented along the $z$ axis. In particular, $\bm m_A=\expval{\bm{s}_A}_{\rm MF} = B(\xi) \bm{u}_z$ and $\bm m_B=\expval{\bm{s}_B}_{\rm MF} = -B(\xi) \bm{u}_z$. 
    All the details leading to an equation of motion for the control parameter $\bm{w}_\alpha$, are provided in the Appendix~\ref{app:spin-wave-details}, taking into account both the inter and intra-sublattice Gilbert damping parameters.
    In the following, for the sake of clarity, as already done in Sec.~\ref{sec:2-time correlators}, we omit the inter-sublattice parameters $\lambda'$ and $\delta \lambda'$, whose presence does not modify the structure of the equations of motion. Similarly to the equation of motion describing spin-spin correlation functions, a decoupling occurs in the dynamics along the direction aligned with the staggered magnetization and within the plane perpendicular to it. In the following, we focus on temperatures that are below the N\'eel temperature.

    \subsubsection{Out-of-plane dynamics}

    The equations of motion for the out-of-plane coordinates of $\bm{w}_\alpha$ are the following
    \begin{equation}
    \label{eq:wz}
    \begin{aligned}
    \dot{w}_{\alpha z} &= -\Theta_{z} w_{\alpha z} -  \Gamma_z' w _{\bar \alpha z} + \kappa _z~,
    \end{aligned}
    \end{equation}
    %
    where $\alpha\in\{A,B\}$,
    \begin{subequations}
    \begin{align}
    \Theta_{z} 
     &= 2\lambda k_{\rm B} T\bigg\{1-\frac{3 k_{\rm B} T+(J+\Delta)[B^2(\xi)-1]}{2 k_{\rm B} T+(J+\Delta)[B^2(\xi)-1]}\bigg\}~,
\\
    \Gamma_{z} '&= 2 \lambda k_{\rm B} T \label{eq:Gammapz}~,\\
    \kappa_z &
    = -\frac{2 \lambda k_{\rm B} T }{2 k_{\rm B} T+(J+\Delta)[B^2(\xi)-1]}h^{\rm ext}_z~. \label{eq:kappaz}
    &
    \end{align}
    \end{subequations}
    %
    We recall that $\xi$ is defined as a function of $k_{\rm B} T/(J+\Delta)$, as described in Eq.~\eqref{eq:self}. 
    For temperatures approaching zero,
    one has $\Theta_z \approx 2 \lambda [k_{\rm B} T +(J+\Delta)]$ and $\kappa_z \approx 2 \lambda \big(\frac{J+\Delta}{k_{\rm B} T}-1\big)  h^{\rm ext}_z$. 
    For $T\to T_{\rm N}^-$, one has $\Theta_z \approx 2 \lambda k_{\rm B} T $ and $\kappa_z \approx 2 \lambda h^{\rm ext}_z$.
%
    
    Assuming a monochromatic external field
    \begin{equation}
    h^{\rm ext}_z(t)=\bar{h}^{\rm ext}_{z}\cos(\Omega_{\rm ext} t)~,
    \end{equation}
    Eq.~\eqref{eq:kappaz} yields $\kappa_z(t)=\bar{\kappa}_z\cos(\Omega_{\rm ext} t)$. By neglecting any transient, the solution of Eq.~\eqref{eq:wz} is
    \begin{equation}
    w_{\alpha z}(t)=\frac{\Theta_\parallel \cos(\Omega_{\rm ext} t) + \Omega_{\rm ext} \sin(\Omega_{\rm ext} t)}{\Theta_\parallel^2+\Omega_{\rm ext}^2} \tilde{\kappa }_{z}~,
    \end{equation}
    where
    \begin{equation}
    \Theta_\parallel=\Theta_{z} + \Gamma_{z}'=2\lambda k_{\rm B} T\bigg\{2-\frac{3 k_{\rm B} T+(J+\Delta)[B^2(\xi)-1]}{2 k_{\rm B} T+(J+\Delta)[B^2(\xi)-1]}\bigg\}~.
    \end{equation}
    At both the limiting values of the temperature interval,
    one has $\Theta_\parallel \approx \Gamma_\parallel$, which is a characteristic damping rate of the spin-spin correlation function involving out-of-plane spin coordinates, as seen in Eq.~\eqref{eq:out-of-plane 2-time correlator}.
    The response $w_{\alpha z}(t)$ associated with an arbitrary $\kappa_z(t)$ is shown in Appendix~\ref{app:spin-wave-details}.

    \subsubsection{In-plane dynamics}
    
    The following equations of motion regard the in-plane coordinates of $\bm{w}_\alpha$
    \begin{equation}\label{eq:dotwbot}
    \dot{\bm{w}}_\bot={\cal A}_\bot \bm{w}_{\bot}+\bm{\kappa}_{\bot}~,
    \end{equation}
    where  $\bm{w}_\bot=(w_{Ax},w_{Bx},w_{Ay},w_{By})^{\rm T}$, 
    \begin{equation}
    \begin{aligned}
    \bm{\kappa}_{\bot}&=
    \frac{J+\Delta}{k_{\rm B} T}\Bigg[B(\xi)(-h^{\rm ext}_y, h^{\rm ext}_y, h^{\rm ext}_x, -h^{\rm ext}_x)\\
    &+(\lambda+\delta \lambda)\bigg(1-\frac{k_{\rm B} T}{J+\Delta}\bigg) (h^{\rm ext}_x, h^{\rm ext}_x, h^{\rm ext}_y, h^{\rm ext}_y)\Big]^{\rm T}~,
    \end{aligned}            
    \end{equation}
    and the expression for the matrix ${\cal A}_\bot$ is given in blockwise form as
    \begin{subequations}
    \begin{align}
    {\cal A}_\bot&=
    \begin{pmatrix}
    {\cal M} & {\cal M}'\\
    -{\cal M}' & {\cal M}    
    \end{pmatrix}~,\\
     {\cal M}&=
    \begin{pmatrix}
   -\Theta_\bot &-\Gamma'_\bot \\
     -\Gamma'_\bot&-\Theta_\bot
    \end{pmatrix}~,\\
    {\cal M}'&=B(\xi)
    \begin{pmatrix}
    J+\Delta & J\\
    -J & -(J+\Delta)  \label{eq:Mp}\\
    \end{pmatrix}~,
    \end{align}
    \end{subequations}
    where $\Gamma_\bot'$ has been defined in Eq.~\eqref{eq:Gammapbot}, and 
    \begin{equation}
    \Theta_\bot=(\lambda+\delta \lambda)(J+\Delta)-(\lambda+2\delta \lambda)k_{\rm B}T~.
    \end{equation}
    Moreover, close to both zero temperature and nearly below the   N\'eel temperature, $T_{\rm N}$, one has $\Theta_\bot \approx \Gamma_\bot$, which is the characteristic damping rate of the spin-spin correlation function involving in-plane spin coordinates, as seen in Eq.~\eqref{eq:in-plane dynamics 2-time correlators}. 
    
    Analogously to the out-of-plane scenario, we assume that the external field is monochromatic 
    \begin{equation}
    h^{\rm ext}_i(t)=\bar{h}^{\rm ext}_{i}\cos(\Omega_{\rm ext} t)~,
    \end{equation}
    with $i\in \{x,y\}$, so that $\bm{\kappa}_\bot(t)=\bar{\bm{\kappa}}_\bot \cos(\Omega_{\rm ext}t)$.  By neglecting any transient, the solution of Eq.~\eqref{eq:dotwbot} is 
    \begin{equation}\label{eq:responsebot}
    \bm{w}_\bot=2 \Re\Big[ \frac{(\Theta_\bot + i \Omega)\cos(\Omega_{\rm ext}t)+\Omega_{\rm ext} \sin(\Omega_{\rm ext}t)}{\Omega_{\rm ext}^2-(\Omega-i \Theta_\bot)^2} \Xi \tilde{\bm{\kappa}} \Big]~,
    \end{equation}
    where $\Omega$ has been defined in Eq.~\eqref{eq:properOmega}, and the matrix $\Xi$ given in blockwise form as

    \begin{subequations}
    \begin{align}
    \Xi&=\frac{i}{\Omega}
    \begin{pmatrix}
    {\cal N} & {\cal M}'\\
    -{\cal M}' & {\cal N}    
    \end{pmatrix}~,\\
     {\cal N}&=
    \begin{pmatrix}
   -i \Omega &-\Gamma'_\bot \\
     -\Gamma'_\bot&-i \Omega
    \end{pmatrix}~,
    \end{align}
    \end{subequations}
    and $\cal M'$ is defined in Eq.~\eqref{eq:Mp}.
    The expression for $\bm{w}_\bot$ in Eq.~\eqref{eq:responsebot} shows that a resonance can be excited at the characteristic energy $\Omega$, while $\Theta_\bot$ determines the resonance linewidth.
    The response ${\bm w}_{\bot}(t)$ to a generic $\bm{\kappa}_\bot(t)$ is discussed in Appendix~\ref{app:spin-wave-details}.
    In conclusion, we note that the adopted methodology incorporates corrections for thermal fluctuations that affect the characteristic energy of spin-waves and the associated decay rates, beyond the MF approximation ~\cite{kamra_prb_2018}. Specifically, in the absence of corrections arising from thermal fluctuations, it is sufficient to replace $\Theta_\bot$ with $(\lambda+\delta \lambda) (J+\Delta) B^2(\xi)$ and to set $\Gamma_\bot' = 0 $.
In the zero-temperature limit, $T \to 0$, the resonance frequency and its corresponding decay rate coincide with those reported in Ref.~\cite{kamra_prb_2018}.

    \subsection{Thermal fluctuations in the resistance}
    \label{sec:PSD}

\begin{figure}[t]
            \centering
            \includegraphics[width=\columnwidth]{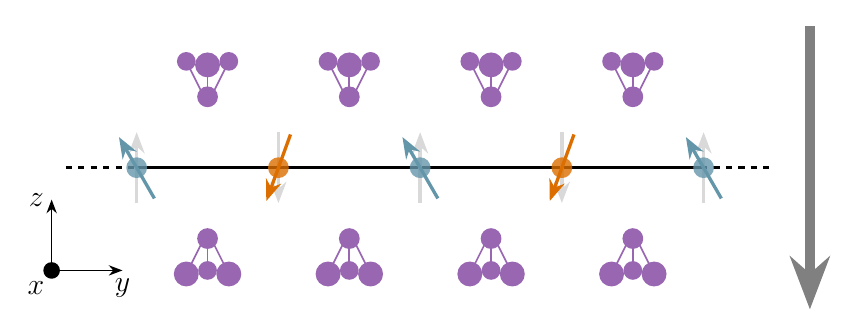}
            \caption{
            Side view sketch of the 2D antiferromagnetically-ordered systems MPX\textsubscript{3}. Vertical electronic transport (see the large gray arrow on the right side of the schematics) occurs principally as a result of hopping between atomic groups (purple structures) located outside the plane.
            The orange and blue dots represent the AFM planar system, which consists of two sublattices, each with a spin on each site.
            Within the macrospin approximation, the deviation from the exact AFM order is represented here, and it generates an internal stray field. The exact AFM pattern is represented by light gray arrows, in which the spin directions of the sublattices $A$ and $B$ are antiparallel.
            }
            \label{fig:MPX3 sketch}
        \end{figure}

        As mentioned in the introduction, this work is motivated by the experimental observations presented in~\cite{ghosh2024electronic}, where vertical resistance fluctuations in a 2D antiferromagnetic semiconductor showed an anomalous behavior close below the N\'eel temperature. 
        The compound used in the experiment is a transition metal-phosphorus-trichalcogenides (MPX\textsubscript{3})~\cite{chittari2016electronic,nitschke2023valence,ghosh2024electronic,samal2021two}. 
        In these compounds, transition metal atoms (M) lay on a plane forming a honeycomb lattice, with negligible orbital overlap that inhibits conduction electrons from hopping within the plane, electron transport, however, takes place principally via the phosphorus-tricalchogenide groups (PX\textsubscript{3}) that take the shape of pyramids above and below the metal layer, and mediate longitudinal charge transport as sketched in Fig.~\ref{fig:MPX3 sketch}~\cite{chittari2016electronic,nitschke2023valence,ghosh2024electronic,samal2021two}. 

    \begin{figure}[t]
	\centering
    \begin{overpic}[width=\columnwidth]{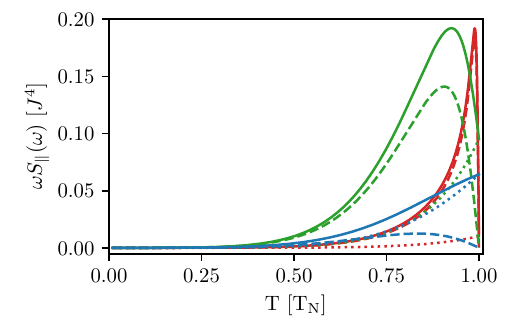}\put(5,55){a)}\end{overpic}\vspace{0em}
	\begin{overpic}[width=\columnwidth]{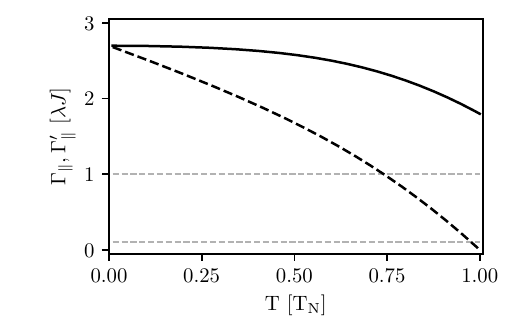}\put(5,55){b)}\end{overpic}\vspace{0em}
        \caption{ 
        a) Out-of-plane term $S_{\parallel}(\omega)$ of the resistance power spectrum as a function of the temperature, ranging from the zero temperature to the N\'eel temperature, for three values of $\omega$: $\omega=\lambda J/10$ (red lines), $\omega=\lambda J$ (green lines), and $\omega=20 \lambda J$ (blue lines). Solid lines represent $S_{\parallel}(\omega)$ in units of $J^4/\omega$, as defined in Eq.~\eqref{eq:Sparallel} where we see that $S_{\parallel}(\omega)$ can be decomposed into a term characterized by a Lorentzian function with a width of $2\Gamma_\parallel$ (dotted lines) and another term (dashed line) that comprises two Lorentzian functions with widths $\Gamma_\parallel'$ and $2\Gamma_\parallel'$.
        b) Damping rates $\Gamma_\parallel$ (black solid line) and $\Gamma_\parallel'$ (black dashed line), defined in Eq.~\eqref{eq:out-damping}, as a function of temperature. Here, the horizontal gray lines correspond respectively to $\lambda J/10$ and $\lambda J$.
        Results in all panels have been obtained by setting  $\Delta=0.35 J$. 
        }
        \label{fig:PSD_z}
        \end{figure}
        
        In this framework, one can approximate the electrons responsible for conduction as distinct from those responsible for the magnetic properties~\cite{YamadaTakada1973,chittari2016electronic,nitschke2023valence}.
        Based on the definition of the internal magnetic field for the sublattice in Eq.~\eqref{eq:eff-h}, and in the macrospin approximation, we can define the net internal magnetic field as the sum of the two fields $\vb*h_A$ and $\vb*h_B$:
        \begin{equation}
        \label{eq:total field}
            \vb*h= -J \sum_{i=x,y}(s_{A i} + s_{B i}) \vb*u_i - (J+\Delta) (s_{A z} + s_{B z}) \vb*u_z~.
        \end{equation}
        In the presence of the exact AFM configuration $\bm{s}_A=-\bm{s}_B$, the conducting electrons are not affected by a net internal magnetic field.
        When a disturbance disrupts the perfect AFM order, a non-zero net internal field can be produced. Using the methods shown in Sec.~\ref{sec:EOM}, we develop a phenomenological model to describe thermal fluctuations as the source of this disturbance.
        %
        %
%
According to the scheme represented in Fig.~\ref{fig:MPX3 sketch}, it is assumed that transport, due to conducting electrons in a (stray) magnetic field, generates an electrical resistance that depends quadratically on the intensity of the internal net magnetic field $\bm{h}$ as
        \begin{equation}
        \label{eq:magnetoresistance}
        R=R_0 +\frac{\kappa_R}{2} h^2+\frac{\kappa_R'}{2}h_z^2~,
        \end{equation}
where $R_0$, $\kappa_R$, and $\kappa_R'$ denote phenomenological parameters. Here, $R_0$ is the resistance in the absence of any interaction with the AFM system, $\kappa_R$ is a proportionality constant that is taken to be independent of the magnetic configuration, and $\kappa_R'$ is a similar parameter that can additionally account for a cylindrical symmetry in this effective magnetoresistance.
The underlying physical picture is that the electrons responsible for charge transport are distinct from those determining the magnetic behavior, similarly to the model for antiferromagnetic metals described in Ref.~\cite{YamadaTakada1973}.
We proceed in analogy with conducting systems subjected to a weak external magnetic field, for which the electrical resistance usually exhibits a quadratic dependence on the intensity of the applied magnetic field~\cite{hu2008classical}, including the cases with vertical tunneling transport~\cite{long_nanolett_2020,huang_nanolett_2023,esaki_prl_1967}.
Accordingly, within the present framework, the resulting internal net magnetic field can affect the electrical resistance.
We therefore treat the influence of the effective net field as a perturbation and assume that the resistance can be systematically expanded in powers of this field. In addition, the 2D system is assumed to be invariant under spatial inversion, as in the minimal description of MPX$_3$ compounds on a honeycomb lattice. As a result, the lowest-order nonzero terms in the perturbative series are quadratic in the effective net field. Finally, because we use a 2D honeycomb lattice as our reference model (with analogous results applying to a 2D square lattice), the system is rotationally invariant about the $z$-axis in the long-wavelength limit~\cite{Dresselhaus_book}. This symmetry reduces the description to only two phenomenological coefficients, $\kappa_R$ and $\kappa_R'$.
For simplicity, we take the resistance to depend spherically on the net effective magnetic field, which is equivalent to setting $\kappa_R' = 0$.

        With the application of the FP equation, the resistance $R(t)$ is considered a function of the random variables $\bm{s}_{\alpha}(t)$, linked to the internal net magnetic field $\bm{h}(t)$.
        It is useful to define the effective magnetoresistance as
         $R(t)=R_0 +  \kappa_R[R_\parallel(t) + R_\bot(t)] $, where
        \begin{subequations}
        \label{eq:resistance contributions}
            \begin{align}
                            R_\parallel(t) &= (J + \Delta)^2 s_{A z}(t)s_{B z}(t)~,\\
                R_\bot(t) &= J^2 \sum_{i =x,y} s_{A i}(t)s_{B i}(t)  \\
                &- \frac{\Delta(2J + \Delta )}{2}\sum_{i =x,y}  [s_{A i}^2(t) +s_{B i}^2(t)]~.  \nonumber
            \end{align}
        \end{subequations}
        Our aim is to determine the two-time correlation of resistance fluctuations, namely $\expval{\delta R(t)\,\delta R(0)}$, where $\delta R(t)=R(t)-\expval{R}$.
        In particular, we consider being in a stationary regime, and employing the results of Sec.~\ref{sec:stationary}, we have the following
        \begin{equation}
            \expval{R}=R_0- \kappa_R \Big[\Delta (2J+\Delta) \frac{B(\xi)}{\xi} +(J+ \Delta)^2 B^2(\xi)\Big]~,
        \end{equation}
        where the terms in square brackets are the contributions in- and out-of-plane, respectively.
        Since we are interested in the resistance correlation function $\expval{\delta R(t)\, \delta R(0)}$, we write  $\delta R(t)= R(t) - \expval{R} =\kappa_R[\delta R_\parallel(t) +  \delta R_\bot(t)]$, where
        \begin{subequations}
        \begin{align}
            \delta R_\parallel(t)&=  (J+\Delta)^2 [s_{A z}(t)s_{B z}(t)+B^2(\xi)]~,\label{eq:dRp}\\
            \delta R_\bot(t)&=\sum_{i=x,y}\Bigg\{ J^2 s_{A i}(t)s_{B i}(t) -  \frac{\Delta (2J + \Delta)}2 \label{eq:dRbot}\\&\times \sum_{\alpha=A,B}\Big[ s_{\alpha i}^2(t) - \frac{B(\xi)}{\xi}\Big]\Bigg\}~. \nonumber 
        \end{align}
        \end{subequations}

      We analyze both the out-of-plane and in-plane contributions. A detailed analysis of these calculations is provided in Appendix~\ref{app:resistance}. Specifically, Appendix~\ref{app:resistance} shows how, by assuming the applicability of Wick's theorem, in the spirit of the Gaussian approximation, we express the two-time correlation function $\expval{\delta R(t) \delta R(0)}$ in terms of the spin-spin correlation functions studied in Sec.~\ref{sec:2-time correlators}.
      The resistance-resistance correlation function related to the out-of-plane magnetic fluctuations is expressed as
      \begin{widetext}
      \begin{equation}\label{eq:W-Rpp}
      \begin{aligned}
      \expval{\delta R_\parallel(t) \delta R_\parallel(0)}&=(J+ \Delta)^4 \{B^2(\xi)[g_{Az,Az}(t)+g_{Bz,Bz}(t)-g_{Az,Bz}(t)-g_{Bz,Az}(t)]\\
      &+g_{Az,Az}(t) g_{Bz,Bz}(t)+g_{Az,Bz}(t) g_{Bz,Az}(t)\}~,
      \end{aligned}
      \end{equation}
       and the one associated with the in-plane magnetic fluctuations is 
       \begin{equation}\label{eq:W-Rbot}
       \begin{aligned}
       \expval{\delta R_\bot(t)\delta R_\bot(0)}&=2\Big\{(J^4+\Omega_0^4)[g^2_{Ax,Ax}(t)+g^2_{Ax,Bx}(t)]    -(J^4-\Omega_0^4)[g^2_{Ax,Ay}(t)+g^2_{Ax,By}(t)]\\
       &-4 \Omega_0^2 J^2 g_{Ax,Ax}(t)g_{Ax,Bx}(t)\Big\},
       \end{aligned}
       \end{equation}
       \end{widetext}
       where $\Omega_0=\sqrt{\Delta(2J+\Delta)}$.

       Our goal is to calculate the resistance power spectrum~\cite{wiener1930generalized,khintchine1934korrelationstheorie,clerk2010introduction}, defined as
       \begin{equation}\label{eq:SR}
       S_{R}(\omega)=\int_{-\infty}^\infty dt e^{i \omega t} \expval{\delta R(t)\delta R(0)}=
       \kappa_R^2 [S_\parallel(\omega)+ S_\bot(\omega)]~,
       \end{equation}
       here, expressed in terms of the Fourier transforms of the correlators, shown in Eqs.~\eqref{eq:W-Rpp} and \eqref{eq:W-Rbot}.
       The out-of-plane contribution is written as
       \begin{equation}\label{eq:Sparallel}
       \begin{aligned}
       S_{\parallel}(\omega)&= \int^{\infty}_{-\infty} dt e^{i \omega t}
        \expval{\delta R_\parallel(t) \delta R_\parallel(0)}=4\pi(J+ \Delta)^4  \sigma_z^2
       \\
        &\times \Big\{  B^2(\xi)  {\cal L}(\omega, \Gamma'_\parallel)
        +
        \frac{\sigma_z^2}{4} \Big[ {\cal L}(\omega, 2\Gamma_\parallel) 
        + {\cal L}(\omega, 2\Gamma_\parallel')\Big] 
        \Big\}~, 
        \end{aligned}
        \end{equation}
        where $\sigma_z$ has been defined in Eq.~\eqref{eq:sigmaz}, and we used the following notation to describe the Lorentzian function
        \begin{equation}\label{eq:Lorentzian}
        \mathcal L (x, y)= \frac 1\pi \frac {y}{x^2+y^2}~.
        \end{equation}
        Using Eq.~\eqref{eq:out-damping}, we note that, if the frequency $\omega$ is expressed in units $\lambda J$, as $\omega/(\lambda J)$, the quantity $S_{\parallel}(\omega)$ can be written as the product of a function of $\omega/(\lambda J)$, independent of $\lambda$, and a factor $1/\lambda$. 
        This allows us to report the results independently of the chosen value of $\lambda$.

        Fig.~\ref{fig:PSD_z}~(a) shows the out-of-plane component $S_\parallel(\omega)$ as a function of the temperature, covering the range from zero to N\'eel temperature, and setting $\Delta/J=0.35$. 
        In particular, we consider three values of $\omega$: $\omega=\lambda J/10$ (solid red line), $\omega=\lambda J$ (solid green line) and $\omega=20 \lambda J$ (solid blue line). Moreover, for each $\omega$, as reference, the dashed line and the dotted line represent, respectively, the contributions in Eq.~\eqref{eq:Sparallel} that involve the damping rate $\Gamma_\parallel'$, which is $\propto  4\sigma_z^2 B^2(\xi)  {\cal L}(\omega, \Gamma'_\parallel)
        +
        \sigma_z^4 {\cal L}(\omega, 2\Gamma_\parallel') $ and $\Gamma_\parallel$, which is $\propto \sigma_z^4 {\cal L}(\omega, 2\Gamma_\parallel)$. Here, we observe that in $S_\parallel(\omega)$, the term involving $\Gamma_\parallel$ (dotted lines) exhibits a monotonic behavior with temperature arising from the increase of thermal fluctuations.
        Specifically, as the temperature increases, there is an enhancement of the prefactor $\sigma_z^4$. 
        On the other hand, the terms of $S_\parallel(\omega)$ involving $\Gamma_\parallel'$ exhibit a peak at a temperature lower than $T_{\rm N}$, and the specific value depends on the chosen $\omega$. 
        In particular, unlike $\sigma_z$, the staggered magnetization decreases with increasing temperature, thus the interplay between $\sigma_z$ and $B(\xi)$ results in the non-monotonic behavior and the formation of a peak.
        Moreover, it has been observed that when $\omega \lesssim \lambda J$, the terms involving $\Gamma_\parallel'$ play a dominant role. As a result, the entire $S_\parallel(\omega)$ displays a distinct peak at a temperature lower than $T_{\rm N}$. 
        %
        %
        Fig.~\ref{fig:PSD_z}~(b) shows the behaviour of the characteristic damping rates $\Gamma_\parallel$ (black solid line) and $\Gamma_\parallel'$ (black dashed line) as functions of temperature. The horizontal lines represent the values of $\omega$: $\lambda J/10$ and $\lambda J$, used in Fig.~\ref{fig:PSD_z}~(a), serving as references.
        Within the explored temperature range, $\Gamma_\parallel'$ vanishes at N\'eel temperature, while $\Gamma_\parallel$ remains finite. This panel provides a quantitative criterion for determining when the term associated with $\Gamma_\parallel$ becomes negligible, that is, for $ \omega < \Gamma_\bot$.

        Analogously, we obtain the following results for the in-plane contribution 
        \begin{widetext}
        \begin{equation}\label{eq:Sbot_ext}
        \begin{aligned}
        S_\bot(\omega)&=\int^{\infty}_{-\infty} e^{i \omega t}
        \expval{\delta R_\bot(t) \delta R_\bot(0)}=2 \pi \sigma_\bot^4 \\
        &\times\bigg\{ (J^4+\Omega_0^4) \bigg\{ \bigg(1+ \frac{\Gamma_\bot'^2}{\Omega^2}\bigg) {\cal L}(\omega,2 \Gamma_\bot)+ \bigg(1- \frac{\Gamma_\bot'^2}{\Omega^2}\bigg) \bigg[ \frac{1}{2}{\cal L}(\omega+2\Omega,2 \Gamma_\bot) +\frac{1}{2}{\cal L}(\omega-2\Omega,2 \Gamma_\bot)\bigg] \bigg\} \\
        &-(J^4-\Omega_0^4)\bigg\{\bigg(\frac{(J+\Delta)^2 +J^2}{\Omega^2}B^2(\xi)\bigg)\bigg[ {\cal L}(\omega,2 \Gamma_\bot)-\frac{1}{2}{\cal L}(\omega+2\Omega,2 \Gamma_\bot) -\frac{1}{2}{\cal L}(\omega-2\Omega,2 \Gamma_\bot)\bigg]\bigg\}\\
        &+4J^2 \Omega_0^2 \frac{\Gamma_\bot'}{\Omega}  {\cal F}_s(\omega,2\Omega,2 \Gamma_\bot) 
        \bigg\}~,
        \end{aligned}
        \end{equation}
        \end{widetext}
        where
        \begin{equation}
        \begin{aligned}
        {\cal F}_s(x,\nu,y)&= \int^{\infty}_{-\infty} e^{i x t} \sin(\nu|t|) e^{-y |t|}\\&=\frac{\nu(y^2+\nu^2-x^2)}{y^4+(\nu^2-x^2)^2+2 y^2(\nu^2+x^2)}~,
        \end{aligned}
        \end{equation}
        and $\sigma_\bot$ has been defined in Eq.~\eqref{eq:sigmabot}. 
              \begin{figure}[t]
                \begin{overpic}[width=\columnwidth]{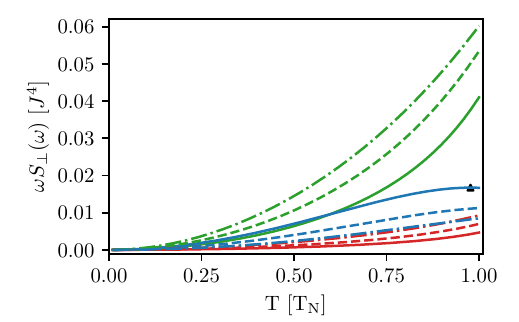}\put(5,55){a)}\end{overpic}\vspace{0em}
                	\begin{overpic}[width=\columnwidth]{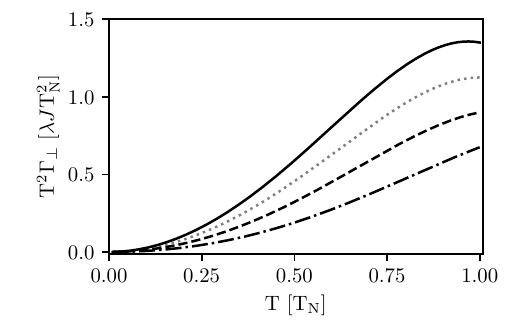}\put(5,55){b)}\end{overpic}\vspace{0em}
                    
                \centering
                \caption{
        a) In-plane term $S_{\bot}(\omega)$ of the resistance power spectrum, defined in Eq.~\eqref{eq:Sbot}, shown as a function of the temperature between the zero temperature and the N\'eel temperature, for different values of $\omega$: $\omega=\lambda J/10$ (red lines), $\omega=\lambda J$ (green lines), and $\omega=20 \lambda J$ (blue lines). 
        Here, the dimensionless parameter $\delta \lambda$ is set $\delta \lambda=\lambda$ (solid lines), $\delta \lambda=0$ (dashed lines), and $\delta \lambda=-\lambda/2$ (dash-dotted lines). We obtain a non monotonic behavior only for $\omega=20 \lambda J$  and $\delta \lambda=\lambda$.
        b) $T^2 \Gamma_\bot$ as a function of temperature, where the damping rate is defined in Eq.~\eqref{eq:Gammabot}. The association between line style and $\delta \lambda$ remains consistent with that in panel~a), while the gray dotted line represents $\delta \lambda=\lambda/2$.
        Results in all panels have been obtained by setting $\Delta=0.35 J$.
        %
        }
                \label{fig:PSD bot}
            \end{figure}    
        Here, we observe that $\Gamma_\bot'/\Omega$ is $\order{{\rm max}(\lambda,\lambda+\delta \lambda)}$, and considering $\lambda, \lambda+\delta \lambda\ll1 $, we can conclude that in Eq.~\eqref{eq:Sbot_ext}, the terms involving the ratio $\Gamma_\bot'/\Omega \ll 1$ are negligible. 
        Moreover, since we are interested in the low-frequency limit, where $\omega \ll \Omega$, in $S_{\bot}(\omega)$, we can discard the contribution of the terms $\mathcal L(\omega \pm 2\Omega, 2 \Gamma_\bot)$
        As a consequence of these remarks, $S_{\bot}(\omega)$  reduces to
        \begin{equation}\label{eq:Sbot}
        \begin{aligned}
        S_\bot(\omega)&=2 \pi \sigma_\bot^4 \bigg\{  (J^4+\Omega_0^4)-(J^4-\Omega_0^4)\\
        &\times\bigg[\frac{(J+\Delta)^2 +J^2}{\Omega_0^2}\bigg] \bigg\}{\cal L}(\omega,2 \Gamma_\bot)~,
        \end{aligned}
        \end{equation}
        which is proportional to the Lorentzian function centered at zero and of width $2 \Gamma_\bot$ multiplied by $\sigma_\bot^4$. Since we are interested in temperatures below $T_{\rm N}$, we replace $\sigma_\bot^4=(k_{\rm B} T)^2/(J+\Delta)^2$. 
        Using Eq.~\eqref{eq:Gammabot}, we see that considering $\omega$ in units of $\lambda J$ makes $S_{\bot}(\omega)$ in the form of a product of $1/\lambda$ and a function of $\omega/(\lambda J)$ and $\delta \lambda/\lambda$. Again, this allows the results to be shown without explicit dependence on $\lambda$.

        Fig.~\ref{fig:PSD bot}~a) shows $ S_\bot(\omega)$ as a function of temperature for three values of $\omega$: 
        $\omega=\lambda J/10$ (red lines), $\omega=\lambda J$ (green lines) and $\omega=20 \lambda J$ (blue lines), 
        while the line style encodes three distinct values of $\delta \lambda / \lambda$: $1$ (solid lines), $0$ (dashed lines) and $-1/2$ (dotted lines).
        
%
%
        Recalling that $\delta \lambda\neq0$ implies that the diffusion coefficients in Eq.~\eqref{eq:DTj} and the Gilbert damping parameters are different, we note that for $\omega \lesssim \lambda J$ when $\delta \lambda>0$, the diffusion term and damping in planar spin-dynamics become more efficient compared to the cases with $-1\le \delta \lambda /\lambda <0$. The lower bound of $\delta \lambda/\lambda=-1$ ensures that diffusive coefficients and the Gilbert damping terms remain non-negative.
        Except for the case with $\omega = 20 \lambda J$ and $\delta \lambda = -1$, which will be discussed below, the in-plane contribution to the resistance power spectrum exhibits a monotonic temperature dependence, as expected. Here, increasing temperature improves fluctuations, which in turn leads to an increase in the prefactor $\sigma_\bot^4 \propto T^2$.

        In Fig.~\ref{fig:PSD bot}~(a), this case is shown by the solid blue line that displays a peak at a temperature slightly below the N\'eel temperature (black triangle). 
        The reason for this behaviour lies in the relation between 
        $\omega $ and $\Gamma_\bot$.
        In particular, if $\omega\gg \Gamma_\bot$ across the entire temperature range examined, it is reasonable to approximate ${\cal L}(\omega,2 \Gamma_\bot) \approx \Gamma_\bot/(\pi \omega^2)$, as a consequence of Eq.~\eqref{eq:Sbot} one has $S_\bot(\omega) \propto T^2 \Gamma_\bot/\omega^2$. 
        The quantity in the numerator is shown in Fig.~\ref{fig:PSD bot}~b), where the different line styles are encoded in the same way as in Fig.~\ref{fig:PSD bot}~a), where we see that for $\delta  \lambda/\lambda=1$ (solid line) $T^2 \Gamma_\bot$ exhibits a peak below $T_{\rm N}$.
        Using Eqs.~\eqref{eq:self} and \eqref{eq:Gammabot}, one finds that as $T\to 0$, the quantity $ T^2 \Gamma_\bot$ exhibits an increasing quadratic temperature dependence, since $\Gamma_\bot \approx (\lambda+\delta \lambda)(J+\Delta)$. 
        On the other hand, as $T\to T_{\rm N}^-$, by using $\xi\propto \sqrt{1-T/T_{\rm N}}$, it shows a linear behavior $T^2 \Gamma_\bot \propto 2+\phi_\bot + (1+3\phi_\bot)\left(\frac{T}{T_{\rm N}}-1\right)$ and is a decreasing function of the temperature for $\phi_\bot < -1/3$. 
        Using Eq.~\eqref{eq:phibot}, it corresponds to $\delta \lambda/\lambda >1/2$.
        The threshold value $\delta \lambda/\lambda =1/2$ is plotted in Fig.~\ref{fig:PSD bot}~(b) as a dotted gray line.
        For values $\delta \lambda >\lambda/2$, the curve of $T^2 \Gamma_\bot$ lies above this limiting case (dotted gray line), becomes non-monotonic, and exhibits a maximum.

            \begin{figure}
                \centering
                \begin{overpic}[width=\columnwidth]{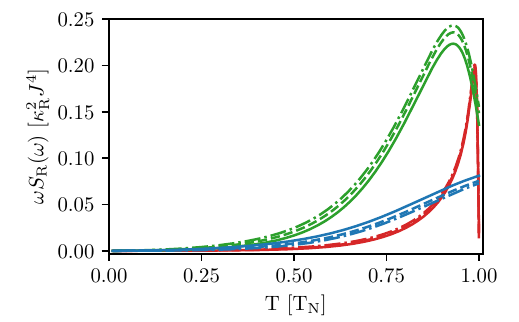} \put(5,55){a)}\end{overpic}\vspace{0em}
                	\begin{overpic}[width=\columnwidth]{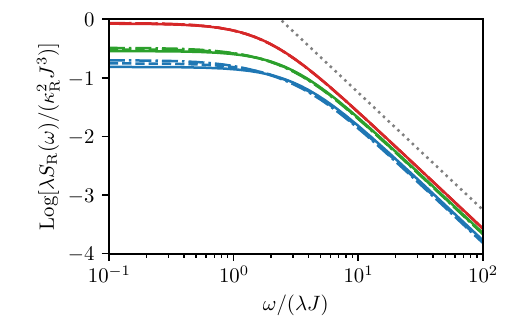}\put(5,55){b)}\end{overpic}\vspace{0em}
                \caption{
                Resistance power spectrum $S_{R}(\omega)$, as defined in Eq.~\eqref{eq:SR}. 
                (a) $S_{R}(\omega)$ as a function of temperature, from zero temperature up to the N\'eel temperature, for different $\omega$: 
                $\omega = \lambda J/10$ (red), $\omega = \lambda J$ (green), and $\omega = 20\lambda J$ (blue). 
                (b) $S_{R}(\omega)$ as a function of frequency for different temperatures: 
                $T = 0.5 T_{\rm N}$ (red), $T = 0.8 T_{\rm N}$ (green), and $T = 0.99 T_{\rm N}$ (blue). 
                In this log-log plot, the gray dotted line indicates a $1/\omega^2$ dependence as a guide to the eye. 
                In both panels, the dimensionless parameter $\delta \lambda/\lambda$ is set to $1$ (solid lines), $0$ (dashed lines), and $-1/2$ (dash-dotted lines). 
                The remaining parameter is $\Delta = 0.35J$.}
                \label{fig:PSD total}
            \end{figure}

        We conclude with a discussion about the full resistance power spectrum.
        Fig.~\ref{fig:PSD total}~a) shows the resistance power spectrum as a function of the temperature for different values of $\omega$ (color-coded) and the dimensionless parameters $\delta \lambda/\lambda$ (line-style-coded).
        We observe that, independently of the value of $\delta \lambda$, and even under the assumption that the resistance depends spherically symmetrically on the net field ${\bm h}$ ($\kappa_R'=0$), the contribution of $S_\parallel(\omega)$ dominates over $S_\bot(\omega)$, as is evident when comparing each curve with their corresponding in Fig.~\ref{fig:PSD_z}~a). Note in particular the maximum below $T_{\rm N}$ for $\omega=\lambda J/10$ and $\omega=\lambda J$. 
        Switching on the cylindrical symmetry in the Gilbert damping factor, $\delta \lambda \neq 0$ slightly modifies the amplitude of the resistance power spectrum.
        Finally, we comment on Fig.~\ref{fig:PSD total}~b), which shows the $\omega$-dependence of the resistance power spectrum for several temperatures (color-coded) and values of $\delta \lambda/\lambda$ (line-style-coded). 
        In this log-log plot, the decreasing trend as $1/\omega^2$ (dotted gray line) for $\omega\gtrsim \lambda J$ is expected, since $S_R(\omega)$ is the sum of Lorentzian functions.

        
        The behavior described by this phenomenological model qualitatively resembles that reported in~\cite{ghosh2024electronic}, where the authors observed an unusual feature in the noise resistance data obtained from measurements in a 2D AFM system, superimposed on the expected $1/f$ noise~\cite{paladino20141,pellegrino_apl_2023,pellegrino_jstat_2019,pellegrino_commphys_2020,hardner_prb_1993}. Specifically, below the N\'eel temperature, the resistance power spectrum exhibits a contribution that can be described by a Lorentzian function with a characteristic damping rate $\Gamma \gtrsim 10$–$10^2~{\rm Hz}$, corresponding to the typical range of low-frequency resistance noise measurements, \eg $\omega=10$~Hz. 
        This feature is most pronounced slightly below the N\'eel temperature, resembling the red and green curves in Fig.~\ref{fig:PSD total}~a).

\section{Conclusions}
\label{sec:conclusions}

In this work, we have presented a FP formalism to describe the impact of thermal noise on uniaxial anisotropic antiferromagnetic systems.
Starting from a stochastic LLG equation, we derived the FP equation governing the probability distribution of the spin configuration, allowing us to obtain analytical expressions for the time evolution of the spin polarization and of the two-time spin-spin correlation functions. Within the mean-field approximation, the framework provides a unified treatment of precessional, damping, and diffusive effects in the presence of thermal noise.

We further applied this formalism to model spin-wave dynamics, finding a renormalization of the characteristic energies and the damping rates due to the thermal fluctuations. 
Moreover, we have defined a phenomenological model that describes the influence of magnetic fluctuations on the resistance noise of 2D antiferromagnetic semiconductors.
We derived analytic expressions for the resistance power spectrum, showing that thermal fluctuations generate Lorentzian noise contributions with characteristic damping rates. 
This model exhibits key qualitative characteristics that resemble the measurements observed experimentally~\cite{ghosh2024electronic}, in particular, the occurrence of Lorentzian low-frequency resistance noise maximized near the AFM transition.

Overall, our approach provides a microscopic stochastic description linking spin dynamics, thermal fluctuations, and transport noise in two-dimensional antiferromagnets. Beyond the specific case analyzed here, this framework can be generalized to include external driving fields or noncollinear magnetic configurations.

\section{Acknowledgments}
The authors thank G.G.N. Angilella, A. Balandin, J. O. Brown, S. Ghosh, and F. Kargar, for fruitful comments on various stages of this work. 
E.M. and E.P. thank the PNRR MUR project PE0000023-NQSTI.
E.P. acknowledges support from COST Action CA21144 superqumap. 
F.P. acknowledges support from the project PRIN 2022 - 2022XK5CPX (PE3) SoS-QuBa - ``Solid State Quantum Batteries: Characterization and Optimization". 
G.F. thanks for the support ICSC - Centro Nazionale di Ricerca in High-Performance Computing, Big Data and Quantum Computing under project E63C22001000006 and Universit\`a degli Studi di Catania, Piano di Incentivi per la Ricerca di Ateneo, project TCMQI. 

\appendix
\section{Kramers-Moyal expansion and the Fokker-Planck equation}
\label{app:lambda'}


    Within this Appendix, we start from Eq.~\eqref{eq:motion}, which represents a Langevin equation. The corresponding stochastic process is  Markovian, and its probability distribution function (PDF) satisfies a master equation, that we derive here in the Kramers–Moyal form~\cite{kramers1940brownian,moyal1949quantum,garcia2007introduction}. In general, the Kramers–Moyal expansion transforms a master equation into infinite-order differential equation.
    When truncated at the second-order, the resulting partial differential equation is known as the Fokker–Planck (FP) equation. In the present case, by employing \eqref{eq:zeta} of the main-text, the truncation is exact~\cite {garcia2007introduction}, and the Kramers-Moyal coefficients are expressed as
    \begin{equation}
        \begin{aligned}
            a^{(1)}_{\alpha i}&={\mathcal A}_{\alpha i}+
            \sum_{\beta j, \eta k} D_{ \mathrm T, k} \frac{\partial {\mathcal B}_{\alpha i, \eta k}}{\partial \mathcal{S}_{\beta j}} {\mathcal B}_{\beta j, \eta k}~,\\
            a^{(2)}_{\alpha i,\beta j}&=2   \sum_{\eta k}  D_{ \mathrm T,k} {\cal B}_{\alpha i, \eta k} {\cal B}_{\beta j, \eta k}~.
        \end{aligned}
    \end{equation}
    By a direct substitution of the definition of $\mathcal A$ and $\mathcal B$, in Eq.~\eqref{eq:coeff for KME} of the main-text, one finds
    \begin{equation}\label{app-eq:a1}
        a^{(1)}_{\alpha i}={\mathcal A}_{\alpha i}- 2 D_\mathrm{T} s_{\alpha i} - (1-\delta_{iz}) \delta D_\mathrm{T} s_{\alpha i}~,
    \end{equation}
    and
    \begin{equation}\label{app-eq:a2}
        \begin{aligned}
            a^{(2)}_{\alpha i,\beta j}&=2 \delta_{\alpha,\beta}\big[D_\mathrm{T} (\delta_{ij}-s_{\alpha,i} s_{\alpha,j}) \\
                    &
+\delta D_{\rm T}\sum_{\ell,\ell'}\epsilon_{i\ell z} \epsilon_{j\ell' z} 
s_{\alpha,\ell} s_{\alpha,\ell'}\big]~.
        \end{aligned}
    \end{equation}
    The FP equation for the PDF $\mathcal P$ can then be written as
    \begin{equation}
        \pdv{\mathcal P}{t}=-\sum_{\alpha i}\pdv{}{\mathcal{S}_{\alpha i}} \qty(a^{(1)}_{\alpha i} \mathcal P) + \frac 12 \sum_{\alpha i,\beta j} \pdv[2]{}{\mathcal{S}_{\alpha i}}{\mathcal{S}_{\beta j}} \qty(a^{(2)}_{\alpha i,\beta j} \mathcal P),
    \end{equation}
    and by a direct substitution of Eqs.~\eqref{app-eq:a1} and \eqref{app-eq:a2}, one finally obtains 
        \begin{widetext}
        \begin{equation*}
             \dot{\mathcal P} (\vb*{\mathcal S},t)=- \pdv{\vb* s_\alpha} \cdot \big\{ \vb* s_\alpha \times {\vb* h}_{\alpha} 
             + D_{\mathrm T} {\vb* s}_\alpha \times (\vb* s_\alpha \times \partial_{\vb* s_\alpha})
             + \delta D_{\mathrm T} {\vb* s}_\alpha \times [(\vb* s_\alpha \times \partial_{\vb* s_\alpha})_z \vb*u_z]
             -\Lambda  \vb* s_\alpha \times (\vb* s_\alpha \times {\vb* h}_{\alpha}) 
             -\Lambda' \vb* s_\alpha \times (\vb* s_{\bar \alpha} \times {\vb* h}_{\bar \alpha}) 
             \big\} \mathcal P (\vb*{\mathcal S},t)~,
        \end{equation*}
        \end{widetext}    
    that is the FP equation for the system reported in Eq.~\eqref{eq:FP equation} of the main-text, where we used Einstein’s summation rule for repeated sublattice indices. 
    Using this equation when evaluating the dynamics of the magnetization vector as described in Eq.~\eqref{eq:dynamics magnetization vector} of main-text, one obtains:
    \begin{equation}
    \label{app-eq:dynamics magnetization vector}
        \begin{aligned}
            \dot{\vb*m}_\alpha =& \expval{\vb* s_\alpha \times \vb* h_\alpha} \\
                &- 2 D_{\mathrm T} \vb* m_{\alpha} - 2  \delta D_{\mathrm T} (m_{\alpha x} \vb*u_x +m_{\alpha y} \vb*u_y) \\
                &- \Lambda \expval{\vb* s_\alpha \times(\vb* s_\alpha \times \vb* h_\alpha)}- \Lambda' \expval{\vb* s_\alpha \times(\vb* s_{\bar\alpha} \times \vb* h_{\bar\alpha})}~.
        \end{aligned}   
    \end{equation}
    %
%
     Following Refs.~\cite{kamra_prl_2017,kamra_prb_2018}, we introduce 
    the inter-sublattice Gilbert damping terms via the diagonal matrix $\Lambda'$.

    By applying the mean-field (MF) approximation, 
    the PDF 
    ${\cal P}_{\rm MF}$ is expressed in Eq.~\eqref{eq:PMF} of the main-text, and one has    
    \begin{equation}
    \begin{aligned}
        \expval{s_{\alpha i}}_\mathrm{MF} &= \int \dd[6]\mathcal{S} \, \mathcal P_\mathrm{MF} s_{\alpha i}\\
            &=\int \dd[3]{\bm s}_\alpha \frac{\exp(\vb*\xi_\alpha \cdot \vb*s_\alpha)}{\mathcal Z (\xi_\alpha)} s_{\alpha i}\\
            &=  \pdv {\ln[\mathcal Z(\xi_\alpha)]}{\xi_{\alpha i}}~. 
    \end{aligned}
    \end{equation}
    By defining  $B(\xi) =\mathcal Z'(\xi) / \mathcal Z(\xi)$, where $\mathcal Z'(\xi) = \pdv{\mathcal Z(\xi)}{\xi}$ and $\mathcal Z(\xi)=4\pi \sinh(\xi)/\xi$,
    we obtain the result presented in the main text: $\expval{\vb* s_\alpha }=B(\xi_\alpha) \vb*\xi_\alpha / \xi_\alpha$, see Eq.~\eqref{eq:MFm} of the main-text.
    Similarly, within the MF approximation, we calculate the second moments
    \begin{equation}
     \expval{s_{\alpha i} s_{\beta j}}_{\rm MF}= \pdv {^2\ln[\mathcal Z(\xi_{\rm A})\mathcal Z(\xi_{\rm B})]}{\xi_{\alpha i}\xi_{\beta j}}~.
    \end{equation}
    For the case where $\beta=\alpha$ ($\beta\neq\alpha$), the expression is represented by Eq.~\eqref{eq:ss} (Eq.~\eqref{eq:sbs}) of the main-text. 
    The results shown above are identically obtained by using the following Gaussian PDF, but relaxing the unit norm condition $s_\alpha=1$: 
    \begin{equation}\label{app-eq:PG}
        \mathcal P_\mathrm{G} = 
        \prod_{\alpha = \mathrm{A, B}} \prod_{{j_\alpha} = {x'_\alpha},{y'_\alpha},{z'_\alpha}} \frac{1} {\sqrt{2 \pi \sigma_{j_\alpha}^2}} e^{-\frac{(s_{\alpha,{j_\alpha}}-m_{j_\alpha})^2}{2 \sigma_{j_{\alpha}^2}}} ~,
        \end{equation}
    where the set of indices $x'_\alpha, y'_\alpha$, and $z'_\alpha$ represents the Cartesian components corresponding to the unit vectors ${\bm u}_{x'_\alpha}, {\bm u}_{y'_\alpha}$, and $ {\bm u}_{z'_\alpha}$. Here, we define ${\bm u}_{z'_\alpha}=\bm{\xi}_\alpha/\xi_\alpha$, and ${\bm u}_{x'_\alpha}$ and ${\bm u}_{y'_\alpha}$ are two orthonormal unit vectors perpendicular to ${\bm u}_{z'_\alpha}$. Additionally, the vector $\bm{m}_{\alpha}$ is given by $\bm{m}_{\alpha}=B(\xi_\alpha) {\bm u}_{z'_\alpha}$.
    Finally, setting 
    \begin{subequations}
    \begin{align}
    \sigma_{x'_\alpha}&=\sigma_{y'_\alpha}=\sigma_{\bot _\alpha} =\sqrt{\frac{B(\xi_\alpha)}{\xi_\alpha}}~,\\
    \sigma_{z'_\alpha} &= \sqrt{1-B^2(\xi_\alpha) - 2 \sigma_{\bot _\alpha}^2}~.
    \end{align}
    \end{subequations}
    We find:
    \begin{equation}\label{app-eq:sG}
     \expval{\bm{s}_{\alpha }}_\mathrm{G} = \expval{\bm{s}_{\alpha }}_\mathrm{MF}~,
    \end{equation}
    \begin{equation}\label{app-eq:sG2}
     \expval{\bm{s}_{\alpha } \otimes \bm{s}_{\beta } }_\mathrm{G}=\expval{\bm{s}_{\alpha } \otimes \bm{s}_{\beta }  }_\mathrm{MF}~.
    \end{equation}
    The assumption about the norm can be reasonable when considering a system subject to small fluctuations.

    We now show the evaluation of each term of Eq.~\eqref{app-eq:dynamics magnetization vector} in turn. 
    Starting with the precession term $\expval{\vb*s_\alpha \times {\bm h}_\alpha}$.
    Applying the definition of the effective magnetic field $\bm{h}_\alpha$ as presented in Eq.~\eqref{eq:eff-h} of the main-text, while considering the sublattices as independent, it can be deduced that the precession term transforms into
    \begin{equation}
    \begin{aligned}
        \expval{\vb*s_\alpha \times {\bm h}_\alpha}= \expval{\vb*s_\alpha} \times \expval{\vb*h_{\alpha}}~.
    \end{aligned}
    \end{equation}
    Analogously, the intra-sublattice term is  expressed as
    \begin{equation}
     \Lambda \expval{\vb* s_\alpha \times ( \vb* s_{\alpha} \times \vb* h_{\alpha})} = \Lambda {\expval {{h}_{\alpha j}}}  {\expval{s_{\alpha j} \bm{s}_\alpha  }}
        - \Lambda  \expval{{\bm h}_\alpha}~,
    \end{equation}
     and the inter-sublattice term is expressed as
    \begin{equation}
     \Lambda' \expval{\vb* s_\alpha \times ( \vb* s_{\bar{\alpha}} \times \vb* h_{\bar \alpha})} = \Lambda'   {\expval{h_{\bar{\alpha} j} {s}_{ \alpha j}  }} {\expval {{\bm s}_{\bar{\alpha} }}}
      -\Lambda' {\expval {{s}_{\bar{\alpha} j}}}  {\expval{s_{\alpha j} {\bm{h}}_{ \bar{\alpha}}  }}~.
    \end{equation}     
     Note that we used Einstein’s summation rule on repeated Cartesian coordinates, $j \in \{x,y,z\}$, 
     
     Now, we focus on the stationary limit, where $\dot{\vb* m_\alpha} =0$. In particular, we seek stationary solutions in the form $\bm{\xi}_A=\xi_{ A z} \bm{u}_z$ and $\bm{\xi}_B=\xi_{B z} \bm{u}_z$.
     Based on these assumptions, one has
     \begin{equation}\label{app-eq:lambda_term}
     \Lambda \expval{\vb* s_\alpha \times ( \vb* s_{\alpha} \times \vb* h_{\alpha})} \to \lambda (J+\Delta) m_{\bar{\alpha }z}\frac{2B(\xi_\alpha)}{\xi_{\alpha}}   \bm{u}_z~,
     \end{equation}
    \begin{equation}\label{app-eq:lambdap_term}
     \Lambda' \expval{\vb* s_\alpha \times ( \vb* s_{\bar{\alpha}} \times \vb* h_{\bar \alpha})} \to -\lambda' J m_{\bar{\alpha }z} \frac{2B(\xi_\alpha)}{\xi_{\alpha}}   \bm{u}_z~.
    \end{equation}  
    Since the precession term goes to zero, the stationary version of Eq.~\eqref{app-eq:dynamics magnetization vector} reduces to
    \begin{equation}
     D_{\rm T} m_{\alpha z}=-[\lambda (J+\Delta)-\lambda' J] m_{\bar{\alpha} z} \frac{B(\xi_\alpha)}{\xi_\alpha}~.
    \end{equation}
     Given the conditions $\lambda/\lambda'\ge1$, $J>0$, and $\Delta>0$, together with the AFM order $m_{\alpha z}=-m_{\bar{\alpha} z}$,  one has
    \begin{equation}\label{app-eq:DT-B}
    D_{\rm T}=[\lambda(J+\Delta)-\lambda'J] \frac{B(\xi)}{\xi}~,
    \end{equation}
    where $\xi=|\xi_{\alpha z}|$. 
    Given that the expression matches the thermodynamic MF solution $\frac{B(\xi)}{\xi}=\frac{k_{\rm B} T}{J+\Delta}$, it follows that
    \begin{equation}\label{app-eq:DT}
    D_\mathrm{T} = \qty(\lambda- \lambda' \frac J{J+\Delta})k_\mathrm{B} T~.
    \end{equation}
    Through a comparison with the case with $\lambda'=0$ as shown in Eq.~\eqref{eq:DTthermal} of the main-text, here we find that the parameter $\lambda'$ renormalizes the proportionality constant of $D_{\rm T}$ in relation to the thermal energy, namely $\lambda\to\lambda\{1-\lambda' J/[\lambda(J+\Delta)]\}$.

    Reiterating these calculations and assuming that the stationary solutions take the form $\bm{\xi}_A=\xi \bm{u}_x$ and $\bm{\xi}_B=-\xi \bm{u}_x$, one has
    \begin{equation}
    D_{\rm T}+\delta D_{\rm T}=[(\lambda+\delta \lambda) J-(\lambda'+\delta \lambda')(J+\Delta/2)] \frac{B(\xi)}{\xi}~,
    \end{equation}
    from a comparison with the local minimum of the free energy via the classical Hamiltonian, using the MF method~\cite{goldenfeld_book}, given by $\frac{B(\xi)}{\xi}=\frac{k_{\rm B} T}{J}$, one has
    \begin{equation}
    D_{\rm T}+\delta D_{\rm T}=\Big[(\lambda+\delta \lambda)  -(\lambda'+\delta \lambda')\frac{J+\Delta/2}{J}\Big] k_{\rm B} T~.
    \end{equation}
    Using Eq.~\eqref{app-eq:DT}, one can extract $\delta D_{\rm T}$ as
    \begin{subequations}
    \begin{align}
    \delta D_{\rm T}&=\qty( \lambda-  \lambda' \frac{J}{J+\Delta})\nu_{\rm T} k_\mathrm{B} T \label{app-eq:dDT-A}\\
    \nu_{\rm T}&=-\frac{J+\Delta}{\lambda(J+\Delta)-\lambda' J}\Big[ 
    \lambda'\frac{(J+\Delta/2)(J+\Delta)-J^2}{J(J+\Delta)}
    \nonumber\\
    &+\delta \lambda' \frac{J+\Delta/2}{J}-\delta \lambda
    \Big]~,
    \end{align}
    \label{app-eq:dDT-B}
    \end{subequations}
    In order to fulfill condition $D_{\rm T}+\delta D_{\rm T}>0$, one has the constraint $\nu_{\rm T}>-1$.
    In the particular scenario where inter-sublattice terms are disregarded,  
    one has $\nu_{\rm T}=\delta \lambda/\lambda$, and the aforementioned constraint is satisfied  
    provided that $\delta \lambda>-\lambda$.

\section{Equation of motion of  correlation functions}
    \label{app:two-time correlators}

    In this Appendix, we derive the equation of motion of the spin-spin correlation functions studied in Sec.~\ref{sec:2-time correlators} of the main-text. For this aim, it is useful to define the following correlation functions
    \begin{widetext}
    \begin{equation}\label{app-eq:g}
    \begin{aligned}
    g_{\alpha_1 i_1\ldots\alpha_N i_N,\beta_1 \ell_1\ldots \beta_M \ell_M}(t)
     &= \int \dd[6]{\vb*{\mathcal S}} \int \dd[6]{\vb*{\mathcal S}'} [s_{\alpha_1 i_1}-m_{\alpha_1 i_1}] \ldots [s_{\alpha_N i_N}-m_{\alpha_N i_N}] [s'_{\beta_1 \ell_1}-m_{\beta_1 \ell_1}]\ldots [s'_{\beta_M i_M}-m_{\beta_M i_M}] \\
     &\times {\cal P}(\vb*{\mathcal S},t|\vb*{\mathcal S}',0){\cal P}_0(\vb*{\mathcal S}')~,
     \end{aligned}
     \end{equation}    
    \end{widetext}
     where  $\mathcal{P}_0(\vb*{\mathcal S})$ represents the probability distribution in the stationary regime~\cite{likharev2019statistical}.
     In the case with $N=1$ and $M=1$, one finds the correlation functions defined with Eq.~\eqref{eq:2-time correlator base} of the main-text.

    Starting with the time derivative of the definition of the correlation functions in Eq.~\eqref{eq:2-time correlator base} of the main-text, we have
    \begin{equation}
    \label{app-eq:dtg}
    \begin{aligned}
        \dot {g}_{\alpha i,\beta \ell}(t) &=\int \dd[6]{\vb*{\mathcal S}} \int \dd[6]{\vb*{\mathcal S}'} s_{\alpha i} [s'_{\beta \ell}-m_{\beta \ell}]  \dot{{\cal P}}(\vb*{\mathcal S},t|\vb*{\mathcal S}',0) \\&\times{\cal P}_0(\vb*{\mathcal S}')~.
    \end{aligned}
    \end{equation}
    By using the FP equation on ${\mathcal P}(\vb*{\mathcal S},t|\vb*{\mathcal S}',0)$, the resulting equation of motion for the correlation function is obtained~\cite{garcia2007introduction,garanin1997fokker}. 
    %
    %
    This equation has the same structure of the equation for the dynamics of the magnetization, with the precession term, the diffusion term, and finally the intra-sublattice and inter-sublattice terms
    \begin{equation}
    \dot {g}_{\alpha i,\beta \ell}(t) =\sum_{r=1}^4 {\cal F}^{(r)}_{\alpha i, \beta j}(t)~,
    \end{equation}
    that we discuss here in turn.
    The precession term is written as
    \begin{equation}
    \begin{aligned}
        {\cal F}^{(1)}_{\alpha i, \beta j} (t)&=- \int \dd[6]{\mathcal{\vb{S}}}\int \dd[6]{\mathcal{\vb{S}'}} s_{\alpha i}  [s'_{\beta j}-m_{\beta j}] \\
        &\times \sum_{\gamma=A,B} \pdv{}{\vb*s_\gamma} \cdot  \vb*s_\gamma \times \vb* h_\gamma  {\mathcal P} (\vb*{\mathcal S}, t | \vb*{\mathcal S}', 0)  \mathcal P_0(\vb*{\mathcal S}')~.
    \end{aligned}
    \end{equation}
    Integrating this by parts,
    this can be rewritten componentwise, using Einstein's summation rule on the repeated indices, and one has
    \begin{equation}
    \label{app-eq:F1}
    \begin{aligned}
         {\cal F}^{(1)}_{\alpha i, \beta j} (t)
         &\to -\epsilon_{ijk} \bar{J}_k \expval{s_{\alpha j}(t) s_{\bar \alpha k}(t) [s'_{\beta \ell}(0)-m_{\beta \ell}]} \\
         &=-\epsilon_{ijk } [\bar{J}_k m_{\alpha j} g_{\bar{\alpha}k, \beta \ell} (t) + \bar{J}_k m_{\bar{\alpha} k} g_{\alpha j, \beta\ell} (t)]\\
         &-\epsilon_{ijk} \bar{J}_k g_{\alpha j \bar{\alpha} k, \beta \ell}(t)~,
    \end{aligned}
    \end{equation}
    where $\bar{J}_k=J+\Delta\delta_{kz}$,
    Analogously, we derive the remaining terms of the equation of motion.
    The diffusion term is expressed as
    \begin{equation} \label{app-eq:F2}
    {\cal F}^{(2)}_{\alpha i, \beta j} (t)=-2D_{\rm T} g_{\alpha i,\beta \ell}(t) -\delta D_{\rm T}(1-\delta_{iz}) g_{\alpha i,\beta \ell}(t)~. 
    \end{equation}
    The intra-sublattice  term is written as
    \begin{equation}
    \begin{aligned} \label{app-eq:F3}
    {\cal F}^{(3)}_{\alpha i, \beta j} (t)&=\Lambda_{ii}\langle[\bar{J}_k s_{\alpha k}(t) s_{\bar{\alpha} k}(t) s_{\alpha i}(t)-\bar{J}_i s_{\bar{\alpha} i}(t)  ]\\&\times[s_{\beta \ell} (0)- m_{\beta \ell}]\rangle\\
    &=\Lambda_{ii}\{  \bar{J}_k [m_{\alpha k} m_{\bar{\alpha} k} g_{\alpha i, \beta \ell}(t)
    + m_{\alpha k} m_{\alpha i} g_{\bar{\alpha} k, \beta \ell}(t)\\
    &+m_{\bar{\alpha} k}  m_{\alpha i} g_{\alpha k, \beta \ell}(t) + m_{\alpha k} g_{\bar{\alpha} k \alpha i, \beta \ell }(t)\\
    & + m_{\bar{\alpha}k} g_{\alpha  k \alpha i, \beta \ell }(t)+ m_{\alpha i} g_{\alpha k\bar{\alpha} k , \beta \ell }(t) \\
    &+g_{\alpha k\bar{\alpha} k \alpha i, \beta \ell }(t)]-  \bar{J}_i g_{\bar{\alpha} i, \beta \ell}(t)\}~,
    \end{aligned}
    \end{equation}
    and the inter-sublattice term is expressed as
    \begin{equation}
    \begin{aligned} \label{app-eq:F4}
    {\cal F}^{(4)}_{\alpha i, \beta j} (t)&=-\Lambda_{ii}'\langle[\bar{J}_i s_{\alpha k}(t) s_{\bar{\alpha} k}(t) s_{\alpha i}(t) -(J +\Delta s_{\alpha z}^2(t))\\
    &\times s_{\bar{\alpha} i}(t)][s_{\beta \ell}(0)-m_{\beta \ell}]\rangle\\
    &=-\Lambda_{ii}'\{  \bar{J}_i [m_{\alpha k} m_{\bar{\alpha} k} g_{\alpha i, \beta \ell}(t)
    + m_{\alpha k} m_{\alpha i} g_{\bar{\alpha} k, \beta \ell}(t)\\
    &+m_{\bar{\alpha} k}  m_{\alpha i} g_{\alpha k, \beta \ell}(t) + m_{\alpha k} g_{\bar{\alpha} k \alpha i, \beta \ell }(t)\\
    & + m_{\bar{\alpha}k} g_{\alpha  k \alpha i, \beta \ell }(t)+ m_{\alpha i} g_{\alpha k\bar{\alpha} k , \beta \ell }(t) \\
    &+g_{\alpha k\bar{\alpha} k \alpha i, \beta \ell }(t)]-(J+\Delta m_{\alpha z}^2) g_{\bar{\alpha} i, \beta \ell}(t)\\
    &-\Delta[2 m_{\alpha z} m_{\bar{\alpha} i}g_{\alpha z,\beta \ell}(t)+ m_{\bar{\alpha} i} g_{\alpha z \alpha z, \beta \ell}(t)\\
    &+2 m_{\alpha z } g_{\alpha z \bar{\alpha} i, \beta \ell}(t)
    + g_{\alpha z \alpha z\bar{\alpha} i, \beta \ell}(t)]\}~.
    \end{aligned}
    \end{equation}    

    The equation of motion for the generic correlation function $g_{\alpha i,\beta \ell}(t)$ involves correlation functions related to higher-order fluctuations, specifically those such as $g_{\alpha_1 i_1 \alpha_2 i_2,\beta \ell}(t)$ and $g_{\alpha_1 i_1 \alpha_2 i_2 \alpha_3 i_3,\beta \ell}(t)$.
    To obtain an exact solution to the equation of motion, it is necessary to solve a hierarchical series of equations. However, this series is infinite, making the task impractical. Subsequently, to obtain a closed form for the correlation functions, the hierarchy must be truncated. To achieve this, we assume the applicability of the Wick's theorem to the correlation functions associated with higher-order fluctuations, allowing them to be expressed in terms of second-order correlation functions.
    In particular, we apply these two assumptions below
    \begin{equation}\label{app-eq:Wick1}
    \begin{aligned}
    g_{\alpha_1 i_1 \alpha_2 i_2,\beta \ell}(t) &\to [\expval{s_{\alpha_1 i_1}(t)}-m_{\alpha_1 i_1}]g_{\alpha_2 i_2,\beta \ell}(t)\\
    &+[\expval{s_{\alpha_2 i_2}(t)}-m_{\alpha_2 i_2}]g_{\alpha_1 i_1,\beta \ell}(t)~,
    \end{aligned}
    \end{equation}
    \begin{equation}\label{app-eq:Wick2}
    \begin{aligned}
    g_{\alpha_1 i_1 \alpha_2 i_2 \alpha_3 i_3,\beta \ell}(t) &\to \expval{[s_{\alpha_1 i_1}(t)-m_{\alpha_1 i_1}][s_{\alpha_2 i_2}(t)-m_{\alpha_2i_2}]}\\
    &\times g_{\alpha_3 i_3,\beta \ell}(t)\\
    &+ \expval{[s_{\alpha_2 i_2}(t)-m_{\alpha_2 i_2}][s_{\alpha_3 i_3}(t)-m_{\alpha_3i_3}]}\\
    &\times g_{\alpha_1 i_1,\beta \ell}(t)\\
    &+ \expval{[s_{\alpha_3 i_3}(t)-m_{\alpha_3 i_3}][s_{\alpha_1 i_1}(t)-m_{\alpha_1i_1}]}\\
    &\times g_{\alpha_2 i_2,\beta \ell}(t)
    ~.
    \end{aligned}
    \end{equation}
    Since we consider the stationary regime, we have $\expval{s_{\alpha i}(t)}=m_{\alpha i}$, which nullifies any $g_{\alpha_1 i_1 \alpha_2 i_2,\beta \ell}(t)$. In addition, we can write $\expval{[s_{\alpha_1 i_1}(t)-m_{\alpha_1 i_1}][s_{\alpha_2 i_2}(t)-m_{\alpha_2i_2}]}=\expval{s_{\alpha_1 i_1}s_{\alpha_2 i_2}}_0-m_{\alpha_1 i_1} m_{\alpha_2i_2}$. 
    Additionally, now we employ the MF approximation outlined in Sec.~\ref{sec:stationary} of the main-text, assuming the configuration to be antiferromagnetic, where ${\bm m}_{\rm A}=-\bm{m}_{\rm B}=B(\xi)\bm{u}_z$, with the staggered magnetization directed along the $z$-axis.
   Based on the prescription above, within the precession term, we retain only the components associated with the second-order fluctuations
    \begin{equation}
    \begin{aligned}
        {\cal F}^{(1)}_{\alpha i, \beta \ell} (t)&\to \epsilon_{ikz} m_{\alpha z} [J  g_{\bar{\alpha} k,\beta \ell}(t)+(J+\Delta)g_{\alpha k,\beta \ell}(t)]~.
    \end{aligned}
    \end{equation}
    The diffusion term is expressed precisely in relation to the correlation function associated with second-order fluctuations, whereas the other terms are approximated by the prescription above.
    The intra-sublattice term appears as
    \begin{equation}
    \begin{aligned} \label{app-eq:F3W}
    {\cal F}^{(3)}_{\alpha i, \beta j} (t)&\to-(\lambda+\delta \lambda)(1-\delta_{iz})[(J+\Delta)m_{\alpha z}^2 g_{\alpha i,\beta \ell}(t)\\
    &+J(1-\sigma_\bot^2) g_{\bar\alpha i,\beta \ell}(t)]\\
    &+\delta_{iz}2\lambda(J+\Delta)   
    [m_{\alpha z}^2 g_{\alpha z, \beta \ell}(t)+\sigma_\bot^2g_{\bar \alpha z, \beta \ell}(t)]
    ~,
    \end{aligned}
    \end{equation}
    and the inter-sublattice term is
    \begin{equation}
    \begin{aligned} \label{app-eq:F4W}
    {\cal F}^{(4)}_{\alpha i, \beta j} (t)&\to-(\lambda'+\delta \lambda')(1-\delta_{iz})[-J m_{\alpha z}^2 g_{\alpha i,\beta \ell}(t)\\
    &-( J(1-\sigma_\bot^2) +\Delta(1-2\sigma_\bot^2) )g_{\bar\alpha i,\beta \ell}(t)]    \\
    &-\delta_{iz}2\lambda' J[m_{\alpha z}^2 g_{\alpha z,\beta \ell}(t)+\sigma_\bot^2 g_{\bar{\alpha} z,\beta \ell}(t)]
    ~,
    \end{aligned}
    \end{equation}
    where $\sigma_\bot=\sqrt{B(\xi)/\xi}$.
    Following the prescription above, we have that the equations of dynamics for $g_{\alpha i,\beta \ell}$ are expressed separately for the in-plane and out-of-plane components, as reported in the main text.

    \subsection{Out-of-plane dynamics}
    \label{ssect:out}
        Before writing the dynamics for the out-of-plane components, it is worth highlighting that the anisotropy terms $\delta D_{\rm T}$, $\delta \lambda$, and $\delta \lambda'$ only affect the in-plane dynamics and that the precession term does not influence the dynamics along the $z$ direction. 
        The equation of motion, which involves the dynamics for the out-of-plane components, is
        \begin{equation}
            \pdv{}{t}
            \begin{pmatrix}
            g_{A z,\beta \ell}(t)\\
            g_{B z,\beta \ell}(t)
            \end{pmatrix}
            =-
            \begin{pmatrix}
            \Gamma_z    &   \Gamma'_z   \\
            \Gamma'_z   &   \Gamma_z    &
            \end{pmatrix}
            \begin{pmatrix}
            g_{A z,\beta \ell}(t)\\
            g_{B z,\beta \ell}(t)
            \end{pmatrix},
        \end{equation}
        where the damping factors are:
        \begin{subequations}
        \label{app-eq:Gammasz}
            \begin{align}
                \Gamma_z &= 2 D_{\rm T}+ 2\qty[\lambda (J+\Delta)-\lambda' J] B^2(\xi)~, \label{app-eq:Gammaz}\\
                \Gamma_z'&= 2 \qty[\lambda (J+\Delta)-\lambda' J]\frac{B(\xi)}{\xi} \label{app-eq:Gammapz}~.
            \end{align}
        \end{subequations}
       Using Eq.~\eqref{app-eq:DT-B}, which gives $D_{\rm T}=\qty[\lambda (J+\Delta)-\lambda' J]B(\xi)/\xi$, and aligning the steady-state solution with the equilibrium thermodynamic solution of the MF such that $B(\xi)/\xi=k_{\rm B} T/(J+\Delta)$, one has the following
         \begin{subequations}
        \label{app-eq:out-of-plane-damping factors}
            \begin{align}
                \Gamma_z &= 2\qty[\lambda (J+\Delta)-\lambda' J] B^2(\xi)+\Gamma_z'~, \\
                \Gamma_z'&= 2 \qty[\lambda (J+\Delta)-\lambda' J]\frac{k_{\rm B} T}{J+\Delta}~.
            \end{align}
        \end{subequations}      
      Finally, applying the initial condition presented in Eq.~\eqref{eq:initial_condition} of the main-text, the solution is expressed as
       \begin{equation}
       \begin{aligned}
       g_{\alpha z, \beta \ell}(t) &= \delta_{\ell z} \Big[1-B^2(\xi)-2\frac{k_{\rm B} T}{J+\Delta}\Big] \Big[ \frac{e^{-(\Gamma_z+\Gamma_z') t}}{2}\\
       &+(2\delta_{\alpha \beta}-1) \frac{e^{-(\Gamma_z-\Gamma_z') t}}{2} \Big]~.
       \end{aligned}
       \end{equation}
        
    \subsection{In-plane dynamics}
    \label{ssect:in-plane}
        Here, we express the equation of motion, which involves the dynamics of the in-plane components, as
        \begin{widetext}
        \begin{equation}
        \label{app-eq:in-plane 2-time dynamics}
            \pdv{}{t}
            \begin{pmatrix}
            g_{Ax,\beta \ell}(t)\\
            g_{Bx,\beta \ell}(t)\\
            g_{Ay,\beta \ell}(t)\\
            g_{By,\beta \ell}(t)
            \end{pmatrix}
            =
            \begin{pmatrix}
            -\Gamma_\bot &-\Gamma'_\bot & (J+\Delta) B(\xi)& JB(\xi)\\
            -\Gamma'_\bot&-\Gamma_\bot & -J B(\xi) & -(J+\Delta) B(\xi)\\
            -(J+\Delta) B(\xi) &-J B(\xi)&-\Gamma_\bot &-\Gamma'_\bot \\
            J B(\xi)& (J+\Delta) B(\xi)&-\Gamma'_\bot&-\Gamma_\bot 
            \end{pmatrix}
            \begin{pmatrix}
            g_{Ax,\beta \ell}(t)\\
            g_{Bx,\beta \ell}(t)\\
            g_{Ay,\beta \ell}(t)\\
            g_{By,\beta \ell}(t)
            \end{pmatrix},
        \end{equation}
        \end{widetext}
        where the damping factors $\Gamma_\bot$ and $\Gamma_\bot'$ are
        \begin{subequations}
        \label{app-eq:Gammasbot}
            \begin{align}
                \Gamma_\bot &= 2 D_{\rm T}+\delta D_{\rm T}  \label{app-eq:Gammabot}\\
                &+\qty[(\lambda+\delta \lambda) (J+\Delta)-(\lambda'+\delta \lambda') J] B^2(\xi)~,\nonumber\\
                \Gamma_\bot'&= \qty[(\lambda+\delta \lambda) J - (\lambda'+\delta \lambda')(J+\Delta)]\bigg(1-\frac{B(\xi)}{\xi}\bigg) \nonumber\\
                &+(\lambda'+\delta \lambda') \Delta \frac{B(\xi)}{\xi} \label{app-eq:Gammapbot}~. 
            \end{align}
        \end{subequations}
         As above, by replacing $B(\xi)/\xi=k_{\rm B} T/(J+\Delta)$, using Eqs.~\eqref{app-eq:DT}-\eqref{app-eq:dDT-B}, we have 
         \newpage
       \begin{subequations}
        \begin{align}
            \Gamma_\bot &=  \qty[\lambda (J+\Delta) -\lambda'J] \qty[(1+\epsilon_{\bot})B^2(\xi) + (2+\nu_{\rm T}) \frac{k_\mathrm{B} T}{J+\Delta} ]~, \label{app-eq:Gammabot_v2}\\
            \Gamma_\bot' &= \qty[\lambda J - \lambda' (J+\Delta)](1+\epsilon'_{\bot})\Big(1-\frac{k_\mathrm{B} T}{J+\Delta} \Big) \nonumber\\
            &+(\lambda'+\delta \lambda') \frac{\Delta }{J+\Delta} k_\mathrm{B} T~, \label{app-eq:Gammapbot_v2}
        \end{align}
        \end{subequations}         
        where 
       \begin{subequations}
        \begin{align}
        \epsilon_\bot&=\frac{\delta \lambda(J+\Delta)-\delta \lambda'J}{\lambda(J+\Delta)-\lambda'J}~,\\
        \epsilon'_\bot&=\frac{\delta \lambda J-\delta \lambda'(J+\Delta)}{\lambda J-\lambda'(J+\Delta)}~.        
        \end{align}
        \end{subequations}  
        In the specific case where inter-sublattice terms $\lambda'$ and $\delta \lambda'$ are ignored, one has $\epsilon_\bot=\delta \lambda/\lambda$ and $\epsilon_\bot'=\delta \lambda/\lambda$, so as $\nu_{\rm T}$ which becomes $\delta \nu_{\rm T}=\delta \lambda/\lambda$. Consequently, neglecting $\lambda'$ and $\delta \lambda'$, the second line of Eq.~\eqref{app-eq:Gammapbot_v2} vanishes.
%
        Applying the initial condition presented in Eq.~\eqref{eq:initial_condition} of the main-text, the solution is expressed as
        \begin{widetext}
        \begin{equation}\label{app-eq:EOM-g-sol}
        \begin{aligned}
        &
        \begin{pmatrix}
        g_{Ax,Ax}(t)&g_{Ax,Bx}(t)&g_{Ax,Ay}(t)&g_{Ax,By}(t)\\
        g_{Bx,Ax}(t)&g_{Bx,Bx}(t)&g_{Bx,Ay}(t)&g_{Bx,By}(t)\\
        g_{Ay,Ax}(t)&g_{Ay,Bx}(t)&g_{Ay,Ay}(t)&g_{Ay,By}(t)\\
        g_{By,Ax}(t)&g_{By,Bx}(t)&g_{By,Ay}(t)&g_{By,By}(t)\\
        \end{pmatrix}=\\
        &=
        \frac{B(\xi)}{\xi} e^{-\Gamma_\bot t}
        \begin{pmatrix}
        \cos(\Omega t)&-\frac{\Gamma_\bot'}{\Omega}\sin(\Omega t)&\frac{(J+\Delta)B(\xi)}{\Omega}\sin(\Omega t)&\frac{J B(\xi)}{\Omega}\sin(\Omega t)\\
        -\frac{\Gamma_\bot'}{\Omega}\sin(\Omega t)&\cos(\Omega t)&-\frac{J B(\xi)}{\Omega}\sin(\Omega t)&-\frac{(J+\Delta)B(\xi)}{\Omega}\sin(\Omega t)\\
        -\frac{(J+\Delta)B(\xi)}{\Omega}\sin(\Omega t)&-\frac{J B(\xi)}{\Omega}\sin(\Omega t)&\cos(\Omega t)&
        -\frac{\Gamma_\bot'}{\Omega}\sin(\Omega t)\\
        \frac{J B(\xi)}{\Omega}\sin(\Omega t)&\frac{(J+\Delta)B(\xi)}{\Omega}\sin(\Omega t)&-\frac{\Gamma_\bot'}{\Omega}\sin(\Omega t)
        &\cos(\Omega t)
        \end{pmatrix}~,
        \end{aligned}
        \end{equation}
        \end{widetext}
        where
        \begin{equation}\label{app-eq:swgap}
        \Omega=\sqrt{\Delta(2J+\Delta)B^2(\xi)-\Gamma_\bot'^2}~.
        \end{equation}    
        %
        %
%

        Note that the fact that $\Gamma_\bot'$ always appears divided by $\Omega$, allows us for further simplification of this expression. In fact, this term only becomes noticeable when $B(\xi) \lesssim \lambda_{\rm max}=\max\{\lambda,\lambda+\delta\lambda,\lambda',\lambda'+\delta \lambda' \}$. Taking into account that $B(\xi)/\xi=k_{\rm B} T/(J+\Delta)$, this corresponds to a very narrow window of temperature, \ie $T_{\rm N}(1-\lambda_{\rm max}^2/15)\lesssim T \le T_{\rm N}$, with $\lambda_{\rm max}\ll 1$. Thus, $\Gamma_\bot'$ leads to minimal influence.

      Until now, our focus has been on $t>0$ concerning the two-time correlation function of the form $\expval{Y(t)Y(0)}$, where $Y$ is a generic random variable. For negative $t$, in a stationary state, we have $\expval{Y(t)Y(0)}=\expval{Y(0)Y(-t)}$. Given that the values of  Y at different times commute, it allows us to rewrite it as $\expval{Y(t)Y(0)}=\expval{Y(-t)Y(0)}$, which is equivalent to $\expval{Y(t)Y(0)}=\expval{Y(|t|)Y(0)}$.

\section{Equation of motion of the spin polarization with a weak external magnetic field}
    \label{app:spin-wave-details}
    In this Appendix, we derive the equation of motion of the spin polarization of each sublattice under a weak external magnetic field perturbing the AFM phase.
    This perturbation is introduced by modifying the magnetic field in Eq.~\eqref{app-eq:dynamics magnetization vector} through the addition of a weak external field $\bm{h}^{\rm ext}$ applied to both sublattices, $\bm{h}_\alpha \to \bm{h}_\alpha+\bm{h}^{\rm ext}$, and 
    using    
    \begin{equation*}
    {\cal P}={\cal P}_{\rm MF} [ 1 + \sum_{\alpha=A,B} \bm{w}_{\alpha} \cdot (\bm{s}_\alpha - \bm{m}_{\alpha}) ]~, 
    \end{equation*}
    %
    %
    where ${\cal P}_{\rm MF}$ is the PDF obtained within the MF approximation, describing the equilibrium configuration, $\bm m_\alpha = \expval{\bm s_\alpha}_{\rm MF}$ denotes the MF spin polarization when no external field is applied, and ${\bm w}_{\alpha}$ is a parameter of order $\mathcal O (h^{\rm ext})$. When the external field is time-dependent, ${\bm w}_{\alpha}$ accordingly acquires a time dependence. The terms of order $\mathcal O ({h^{\rm ext}}^2)$ are neglected. 
    In the remainder of this section, we assume that ${\cal P}_{\rm MF}$ describes the stationary configuration in which the staggered magnetization is aligned along the $z$ axis, namely $\bm m_A = B(\xi) \bm{u}_z$ and $\bm m_B = -B(\xi) \bm{u}_z$. Here $B(\xi)$ is the Langevin function satisfying $B(\xi)/\xi = k_{\rm B} T/(J+\Delta)$, as reported in Appendix~\ref{app:lambda'}.
    
    We now examine each term of Eq.~\eqref{app-eq:dynamics magnetization vector}. The precession term is expressed as
    \begin{equation}
    \begin{aligned}
    \expval{\bm{s}_\alpha \times (\bm{h}_\alpha+\bm{h}^{\rm ext})} &\to    \epsilon_{izk} \Big \{ m_{\alpha z} h_k^{\rm ext}+ \sigma_\bot^2 \\
     &\times\Big [(J+\Delta) m_{\bar{\alpha}z} w_{\alpha k}-J   m_{\alpha z} w_{\bar{\alpha} k}\Big ]\Big \}~,
    \end{aligned}
    \end{equation}
    where $m_{Az}=B(\xi)$ and $m_{Bz}=-B(\xi)$.
    The diffusive term is evaluated using 
    \begin{equation}
    \expval{\bm s_{\alpha i}}= \delta_{iz} m_{\alpha z}+w_{\alpha i} \sigma_i^2~,
     \end{equation}
    where $\sigma_x=\sigma_y=\sigma_\bot=\sqrt{B(\xi)/\xi}$ and $\sigma_z=\sqrt{1-B^2(\xi)-2B(\xi)/\xi}$, and consistently the time derivative of the spin polarization is 
    \begin{equation}
    \begin{aligned}
     \frac{d    \expval{\bm s_{\alpha i}}}{dt}&=\frac{d w_{\alpha i}}{dt} \sigma_i^2~.
     \end{aligned}
    \end{equation}

    We now get to the intra-sublattice interaction term
    \begin{equation}
    \begin{aligned}
    &\Lambda \expval{\bm{s}_\alpha \times [\bm{s}_\alpha \times (\bm{h}_\alpha+\bm{h}^{\rm ext})]}\to   \Lambda \expval{\bm{s}_\alpha \times [\bm{s}_\alpha \times \bm{h}_\alpha]}_{\rm MF}+\\
    &+ 
    \Lambda\expval{\bm{s}_\alpha \times [\bm{s}_\alpha \times \bm{h}^{\rm ext}]}_{\rm MF}\\
    &+\Lambda \expval*{\bm{s}_\alpha \times [\bm{s}_\alpha \times \bm{h}_\alpha]\sum_{\alpha'=A,B}\bm{w}_{\alpha' } \cdot (\bm{s}_{\alpha' } -\bm{m}_{\alpha' })}_{\rm MF}~,
    \end{aligned}
    \end{equation}
    on the right-hand side, the first line coincides with Eq.~\eqref{app-eq:lambda_term}, the second line is 
    \begin{equation}
    \Lambda_{ii} \expval{\bm{s}_\alpha \times [\bm{s}_\alpha \times \bm{h}^{\rm ext}]|_i}_{\rm MF}=\Lambda_{ii} (\expval{s_i^2}_{\rm MF} -1) h^{\rm ext}_i~,
    \end{equation}
    finally, the third line yields the following
    \begin{equation}
    \begin{aligned}
    &\Lambda_{ii} \expval*{\bm{s}_\alpha \times [\bm{s}_\alpha \times \bm{h}_\alpha]|_i\sum_{\alpha'=A,B}\bm{w}_{\alpha' } \cdot (\bm{s}_{\alpha' } -\bm{m}_{\alpha' })}_{\rm MF}=\\
    &=
    -\lambda \delta_{iz} \Big[
    (J+\Delta)\Big(\expval{s_{\alpha z}^3}_{\rm MF}-\expval{s_{\alpha z}^2}_{\rm MF} m_{\alpha z}\Big)m_{\bar{\alpha} z}w_{\alpha z}\\
    &
    -2(J+\Delta)\sigma_\bot^2\sigma_z^2 w_{\bar{\alpha} z}
    \Big]-(\lambda+\delta \lambda) (1-\delta_{iz})\\
    &\times \Big[ 
    (J+\Delta) \expval{s_{\alpha z} s_{\alpha i}^2}_{\rm MF}  m_{\bar \alpha z}w_{\alpha i}
    -J(1-\sigma_\bot^2)\sigma_\bot^2  w_{\bar{\alpha} i}
    \Big]~.
    \end{aligned}
    \end{equation}
    Here, we have to evaluate the following terms 
    \begin{subequations}
    \begin{align}
    \expval{ s_{\alpha i}^2}_{\rm MF}&=\sigma_\bot^2~,\\
    \expval{s_{\alpha z} s_{\alpha i}^2}_{\rm MF}&=\frac{\xi-3B(\xi)}{\xi^2}~, \label{app-eq:szsx2}
    \end{align}
    \end{subequations}
    with $i\neq z$, and using the fact that the vector $\bm{s}_\alpha$ has a unit norm, one has
    \begin{subequations}
    \begin{align}
    \expval{s_{\alpha z}^2}_{\rm MF}&=1-
    \expval{(s_{\alpha x}^2+s_{\alpha y}^2)}_{\rm MF}=1-2\sigma_\bot^2~,\\
    \expval{s_{\alpha z}^3}_{\rm MF}&=\expval{s_{\alpha z}}_{\rm MF}-
    \expval{s_{\alpha z}(s_{\alpha x}^2+s_{\alpha y}^2)}_{\rm MF} \label{app-eq:sz3}\\
    &=(\delta_{\alpha A}-\delta_{\alpha B})\left[B(\xi)-2\frac{\xi-3B(\xi)}{\xi^2}\right]~.\nonumber
    \end{align}
    \end{subequations}
    Similarly, we consider each term of the inter-sublattice term
    \begin{equation}
    \begin{aligned}
    &\Lambda' \expval{\bm{s}_\alpha \times [\bm{s}_{\bar{\alpha}} \times (\bm{h}_{\bar{\alpha}}+\bm{h}^{\rm ext})]}\to   \Lambda' \expval{\bm{s}_\alpha \times [\bm{s}_{\bar{\alpha}} \times \bm{h}_{\bar{\alpha}}]}_{\rm MF}+\\
    &+ 
    \Lambda'\expval{\bm{s}_\alpha \times [\bm{s}_{\bar{\alpha}} \times \bm{h}^{\rm ext}]}_{\rm MF}\\
    &+\Lambda'\expval{\bm{s}_\alpha \times [\bm{s}_{\bar{\alpha}} \times \bm{h}_{\bar{\alpha}}]\bm{w}_{\alpha' } \cdot (\bm{s}_{\alpha' } -\bm{m}_{\alpha' })}_{\rm MF}~,
    \end{aligned}
    \end{equation}    
    on the right-hand side, the first line coincides with Eq.~\eqref{app-eq:lambdap_term}, the second line is 
    \begin{equation}
    \begin{aligned}
    \Lambda'_{ii}\expval{\bm{s}_\alpha \times [\bm{s}_{\bar \alpha} \times \bm{h}^{\rm ext}]|_{i}}_{\rm MF} &=  (\lambda'+\delta \lambda')(\delta_{iz} -1) m_{\alpha z} m_{\bar{\alpha} z}h^{\rm ext}_i\\
    &= (\lambda'+\delta \lambda')B^2(\xi)(1-\delta_{iz} ) h^{\rm ext}_i~,
    \end{aligned}
    \end{equation}
    finally, the third line yields the following
    \begin{equation}
    \begin{aligned}
    &\Lambda_{ii}' \expval*{\bm{s}_\alpha \times [\bm{s}_{\bar{\alpha}} \times \bm{h}_{\bar{\alpha}}]|_i\sum_{\alpha'=A,B} \bm{w}_{\alpha' } \cdot (\bm{s}_{\alpha' } -\bm{m}_{\alpha' })}_{\rm MF}=\\
    &=-\lambda'\delta_{iz} \big[J m_{\bar{\alpha} z} \big(\sum_{j\neq z}\expval{s_{\alpha j}^2 s_{\alpha z}}_{\rm MF} -2\sigma_\bot^2 m_{\alpha z} \big)w_{\alpha z}\\\
    &+2J \sigma_z^2  \sigma_\bot^2 w_{\bar{\alpha} z}
    \big]
    +(\lambda'+\delta \lambda')(1-\delta_{iz})\{
    J  \expval{s_{\alpha z} s_{\alpha i}^2}_{\rm MF}\\
    &\times m_{\bar{\alpha} z} w_{\alpha i}-[(J+\Delta)\expval{s_{\alpha z}^2}_{\rm MF}+
    J \sigma_\bot^2]\sigma_\bot^2 w_{\bar{\alpha} i}
    \}~.
    \end{aligned}
    \end{equation}
     Here, we note that upon substituting $\cal P_{\rm MF}$ with $\cal P_{\rm G}$, defined in Eq.~\eqref{app-eq:PG}, and allowing for relaxation of the unit norm constraint $s_\alpha=1$, it can be observed that the first- and second-order moments remain unchanged, as evidenced by Eqs.~\eqref{app-eq:sG} and~\eqref{app-eq:sG2}. However, there is a modification in the higher-order moments, here more specifically, it pertains to Eqs.~\eqref{app-eq:szsx2} and \eqref{app-eq:sz3}, which yield the results $\expval{s_{\alpha z} s_{\alpha i}^2}_{\rm G}= \expval{s_{\alpha z}}_{\rm G} \expval{s_{\alpha i}^2}_{\rm G}=m_{\alpha z} \sigma_\bot^2$ for $i\neq z$, and $\expval{s_{\alpha z}^3}_{\rm G}=m_{\alpha z}^3+3 m_{\alpha z} \sigma_z^2$.

    Adding up all the calculated terms, the zeroth-order contributions in the external field, which correspond to the stationary solution discussed in Appendix \ref{app:lambda'}, cancel out. Retaining only the terms linear in $\bm{h}^{\rm ext}$, yields a system of linear differential equations, which are intrinsically divided into out-of-plane and in-plane components.
    The out-of-plane equations of motion are the following
    \begin{equation}
    \begin{aligned}
    \dot{w}_{\alpha z} &= -\Theta_{z \alpha} w_{\alpha z} -  \Gamma_z' w _{\bar \alpha z} + \kappa _z~,\\
    \dot{w}_{\bar \alpha z} &=-  \Gamma_z' w_{\alpha z} - \Theta_{\bar \alpha z} w_{\bar \alpha z} + \kappa _z~,
    \end{aligned}
    \end{equation}
    where $ \Gamma_z'$ is defined in Eq.~\eqref{app-eq:Gammapz}, and
    %
    \begin{subequations}
    \begin{align}
    \Theta_{z \alpha} 
     &= [\lambda(J+\Delta)-\lambda'J]\Big\{ 2\sigma_\bot^2\\
     &-\qty[\expval{s^3_{\alpha z} }_{\rm MF} - m_{\alpha z} \expval{s^2_{\alpha z}}_{\rm MF} ]  \frac{m_{\bar\alpha z}}{\sigma^2_z}\Big\}~,\nonumber
\\
    \kappa_z &
    = \lambda \frac{2 \sigma_\bot^2}{\sigma_z^2} h^{\rm ext}_z~.
    \end{align}
    \end{subequations}
%

    Similarly, the equations of motion for the in-plane coordinates are the following
    \begin{equation}\label{app-eq:in-plane_eom}
    \begin{aligned}
      \dot w_{\alpha i} &=
        +\epsilon_{iz \bar i} (J+\Delta) m_{\bar \alpha z} w_{\alpha \bar i} - \epsilon_{iz \bar i} J m_{\alpha z} w_{\bar \alpha \bar i} 
        \\
        &    - \Theta_{\bot \alpha} w_{\alpha i} - \Gamma_\bot' w_{\bar \alpha i} 
        + \kappa_{\alpha i}~,
    \end{aligned}    
    \end{equation}
    where $i=x,y$, $\bar i$ takes the value $x$ ($y$) when $i=y$ ($i=x$), and $ \Gamma_\bot'$ has been defined in Eq.~\eqref{app-eq:Gammapbot}. Using Eqs.~\eqref{app-eq:DT-B} and \eqref{app-eq:dDT-B}, one has
    \begin{subequations}
    \begin{align}
        \Theta_{\bot \alpha } &= 2D_{\mathrm T} +\delta D_{\rm T}
        \\
        &-\Big[(\lambda+\delta \lambda) (J+\Delta) -(\lambda'+\delta \lambda') J\Big] \frac{\expval{s_{\alpha z} s_{\alpha i}^2}_{\rm MF} m_{\bar{\alpha} z}}{\sigma_\bot^2}~,\nonumber \\
     &=  \qty[\lambda (J+\Delta) -\lambda' J] \bigg[ (2+\nu_{\rm T})  \sigma_\bot^2 -(1+\epsilon_{\bot})\nonumber\\
     &\times\frac{\expval{s_{\alpha z} s_{\alpha i}^2}_{\rm MF} m_{\bar{\alpha} z}}{\sigma_\bot^2} \bigg]~,\nonumber\\
        \kappa_{\alpha i} &=\epsilon_{i z \bar{i}} \frac{m_{\alpha z}}{\sigma_\bot^2} h_{\bar{i}}^{\rm ext} + \bigg[(\lambda+\delta \lambda) \frac{1-\sigma_\bot^2} {\sigma^2_\bot}\\
        &+ (\lambda'+\delta \lambda')\frac{m_{\alpha z} m_{\bar \alpha z}} {\sigma^2_\bot}\bigg] h^{\rm ext}_i~. \nonumber
    \end{align}
    \end{subequations}
    %

    We note that $\Theta_{z\alpha}$ and $\Theta_{\bot\alpha}$ do not depend on the sublattice index $\alpha$, since they take the same values for both sublattices. Hence, we can omit the label $\alpha$.
    By treating the third-order moments entering $\Theta_{z}$ and $\Theta_{\bot}$ as Gaussian, namely, recalling that $\expval{s_{\alpha z} s_{\alpha i}^2}_{\mathrm{G}} = \expval{s_{\alpha z}}_{\mathrm{G}} \expval{s_{\alpha i}^2}_{\mathrm{G}} = m_{\alpha z}\sigma_\bot^2$ for $i \neq z$, and $\expval{s_{\alpha z}^3}_{\mathrm{G}} = m_{\alpha z}^3 + 3 m_{\alpha z}\sigma_z^2$, we find that $\Theta_{z}$ and $\Theta_{\bot}$ coincide with $\Gamma_{z}$ and $\Gamma_{\bot}$, defined in Eqs.~\eqref{app-eq:Gammaz} and~\eqref{app-eq:Gammabot}.


    The out-of-plane equations of motion can be expressed compactly as
    \begin{equation*}
    \dot{\bm{w}}_\parallel={\cal A}_\parallel \bm{w}_{\parallel}+\bm{\kappa}_{\parallel}~,
    \end{equation*}
    where  $\bm{w}_\parallel=(w_{Az},w_{Bz})^{\rm T}$, and $\bm{\kappa}_\parallel=(\kappa_{z},\kappa_{z})^{\rm T}$~.
    The matrix ${\cal A}_\parallel$ has a couple of eigenvalues 
    \begin{equation}
    l_\pm=-\Theta_z\pm \Gamma_z'~,
    \end{equation}
    and the corresponding eigenvectors are expressed as
    \begin{equation}
    \bm{\tau}_\pm = (1,\pm1)^{\rm T}~.
    \end{equation}
    The generic vector $\bm{\kappa}_\parallel$ can be expressed as  $\bm{\kappa}_\parallel=\kappa_z \bm{\tau}_+$. 
    So, working in monochromatic components, we define
    \begin{equation}
    \tilde{\kappa}_z(\omega)= \int_{-\infty}^{\infty} dt e^{i \omega t} \kappa_z(t)~,
    \end{equation}
    and the response obtained by the equation of motion is
    \begin{equation}\label{app-eq:wpp}
    \bm{w}_\parallel =\int_{-\infty}^{\infty} \frac{d \omega}{2\pi} e^{-i \omega t} \frac{\tilde{\kappa}_z(\omega)}{-i \omega+\Theta_z+\Gamma_z'} \bm{\tau}_+~.
    \end{equation}

     Similarly, for the in-plane coordinates, one has 
    \begin{equation*}
    \dot{\bm{w}}_\bot={\cal A}_\bot \bm{w}_{\bot}+\bm{\kappa}_{\bot}~,
    \end{equation*}
    where  $\bm{w}_\bot=(w_{Ax},w_{Bx},w_{Ay},w_{By})^{\rm T}$, and $\bm{\kappa}_\bot=(\kappa_{Ax},\kappa_{Bx},\kappa_{Ay},\kappa_{By})^{\rm T}$. 
    In order to analyze the response in $\bm{w}_\bot$ in the presence of $\bm{\kappa}_\bot$, we diagonalize the matrix ${\cal A}_\bot$, and  
    find that it has a pair of two-fold degenerate eigenvalues, namely $\zeta_1=\zeta_2=\zeta$ and  $\zeta_3=\zeta_4=\zeta^\ast$, where
    \begin{equation}
    \zeta=-\Theta_\bot-i\Omega~,
    \end{equation}
    and $\Omega=\sqrt{\Delta(2J+\Delta)B^2(\xi)-\Gamma_\bot'^2}$. 
    The corresponding eigenvectors are expressed as
    \begin{subequations}
    \begin{align}
    \bm{v}_{1}&=(\psi^\ast,\psi^\ast,-1,1)^{\rm T}~,\\ 
    \bm{v}_{2}&=(1,-1,\psi^\ast,\psi^\ast)^{\rm T}~,\\    
    \bm{v}_{3}&=(\psi,\psi,-1,1)^{\rm T}~, \\
    \bm{v}_{4}&=(1,-1,\psi,\psi)^{\rm T}~, 
    \end{align}
    \end{subequations}
    and
    \begin{equation}
    \psi=\frac{-\Gamma_\bot'+i\Omega}{(2J+\Delta)B(\xi)}~. \label{app-eq:psi}
    \end{equation}
    We note that $\bm{v}_3^\ast=\bm{v}_1$ and  $\bm{v}_4^\ast=\bm{v}_2$.
    The generic vector $\bm{\kappa}_\bot$ can be decomposed in terms of the eigenvectors $\bm{v}_r$ as
    \begin{equation}
    \bm{\kappa}_\bot=\sum^4_{r=1} f_r \bm{v}_r~,
    \end{equation}
    where
    \begin{equation}\label{app-eq:fkappa}
    \begin{pmatrix}
    f_1\\
    f_2\\
    f_3\\
    f_4
    \end{pmatrix}
    =
    \frac{i}{2 \Im(\psi)}
    \begin{pmatrix}
    1&1&\psi&-\psi\\
    -\psi&\psi&1&1\\
    -1&-1&-\psi^\ast&\psi^\ast\\
    \psi^\ast&-\psi^\ast&-1&-1
    \end{pmatrix}
    \begin{pmatrix}
    \kappa_{Ax}\\
    \kappa_{Bx}\\
    \kappa_{Ay}\\
    \kappa_{By}
    \end{pmatrix}~.
    \end{equation}
    Also here, we note that, independently of $\bm{\kappa}_\bot$, $f_3=f_1^\ast$ and $f_4=f_2^\ast$.
    Working in monochromatic components, we define
   \begin{equation}
    \tilde{\bm \kappa}_\bot(\omega)= \int_{-\infty}^{\infty} dt e^{i \omega t} {\bm \kappa}_\bot(t)~,
    \end{equation}
    and making use of Eq.~\eqref{app-eq:fkappa}, we have
    \begin{equation}\label{app-eq:tilde_fkappa}
    \tilde{\bm{f}}(\omega)
    =
    \frac{i}{2 \Im(\psi)}
    \begin{pmatrix}
    1&1&\psi&-\psi\\
    -\psi&\psi&1&1\\
    -1&-1&-\psi^\ast&\psi^\ast\\
    \psi^\ast&-\psi^\ast&-1&-1
    \end{pmatrix}
    \tilde{\bm{\kappa}}_\bot(\omega)~,
    \end{equation}
    where we see that
    $\tilde{f}_3(\omega)=\tilde{f}_1(-\omega)^\ast$ and  $\tilde{f}_4(\omega)=\tilde{f}_2(-\omega)^\ast$.
    The response obtained by the equation of motion is
    \begin{equation}
    \bm{w}_\bot =\sum_{r=1}^4\int_{-\infty}^{\infty} \frac{d \omega}{2\pi} e^{-i \omega t} \frac{\tilde{f}_r(\omega)}{-i \omega+\zeta_r} \bm{v}_r~.
    \end{equation}
    Using the properties of $f_r$, $\bm{v}_r$ and $\zeta_r$ ($r=1,\ldots4$), one can write 
    \begin{equation}
    \bm{w}_\bot = \Re \int_{-\infty}^{\infty} \frac{d \omega}{\pi} e^{-i \omega t} \frac{1}{-i \omega+\zeta} [\tilde{f}_1(\omega)\bm{v}_1+\tilde{f}_2(\omega)\bm{v}_2]~,
    \end{equation}
    to write this expression in a more compact way, we define the matrix $\Xi$ as
    \begin{widetext}
     \begin{equation}
     \Xi=    
     \frac{i}{2 \Im(\psi)}
     \begin{pmatrix}
     \bm{v}_1&\bm{v}_2&\bm{0}&\bm{0}
     \end{pmatrix}
    \begin{pmatrix}
    1&1&\psi&-\psi\\
    -\psi&\psi&1&1\\
    -1&-1&-\psi^\ast&\psi^\ast\\
    \psi^\ast&-\psi^\ast&-1&-1
    \end{pmatrix}
    =\frac{i}{2 \Im(\psi)}
    \begin{pmatrix}
    -i2\Im(\psi)&2 \Re(\psi)&1+|\psi|^2&1-|\psi|^2\\
    2 \Re(\psi)&-i2\Im(\psi)&-(1-|\psi|^2)&-(1+|\psi|^2)\\
    -(1+|\psi|^2)&-(1-|\psi|^2)&-i2\Im(\psi)&2 \Re(\psi)\\
    1-|\psi|^2&1+|\psi|^2&2 \Re(\psi)&-i2\Im(\psi)
    \end{pmatrix}~,
     \end{equation}
     that by replacing Eq.~\eqref{app-eq:psi}, it can be written in a recognizable structure
     \begin{equation}
     \Xi         
    =\frac{i}{\Omega}
    \begin{pmatrix}
    -i\Omega &-\Gamma_\bot'&(J+\Delta)B(\xi)&JB(\xi)\\
   -\Gamma_\bot'&-i\Omega&-JB(\xi)&-(J+\Delta)B(\xi)\\
    -(J+\Delta)B(\xi)&-JB(\xi)&-i\Omega& -\Gamma_\bot'\\
   JB(\xi)&(J+\Delta)B(\xi)&-\Gamma_\bot'&-i\Omega
    \end{pmatrix}~,  
     \end{equation}     
    \end{widetext}
    such that 
    \begin{equation}
    [\tilde{f}_1(\omega)\bm{v}_1+\tilde{f}_2(\omega)\bm{v}_2]=\Xi     \tilde{\bm{\kappa}}_\bot(\omega)~,
    \end{equation} 
    and then
   \begin{equation}
    \bm{w}_\bot  = \Re \int_{-\infty}^{\infty} \frac{d \omega}{\pi} e^{-i \omega t} \frac{1}{-i \omega+\zeta} \Xi     \tilde{\bm{\kappa}}_\bot(\omega)~.
    \end{equation}
    %
\section{Resistance-resistance correlation functions}
\label{app:resistance}
    In this Appendix, we provide the detailed calculations performed to derive the results discussed in Sec.~\ref{sec:PSD} of the main-text.
    In particular, we show the underlying assumptions used to represent the two-time correlation function $\expval{\delta R(t) \delta R(0)}$ in terms of the two-time spin-spin correlation functions.
    Starting from the definition of $\expval{\delta R(t) \delta R(0)}$, our task is to calculate correlation functions of the form
    \begin{widetext}
    \begin{equation}\label{app-eq:ss-ss}
    \begin{aligned}
    &\expval{[ s_{\alpha_1 i_1}(t) s_{\alpha_2 i_2}(t)-\expval{s_{\alpha_1 i_1}(t) s_{\alpha_2 i_2}(t)}][ s_{\beta_1 \ell_1}(t) s_{\beta_2\ell_2}(t)-\expval{s_{\beta_1 \ell_1}(t) s_{\beta_2 \ell_2}(t)}]}=\\
    &=m_{\alpha_1 i_1}[ m_{\beta_1 \ell_1} g_{\alpha_2 i_2,\beta_2 \ell_2}(t)+ m_{\beta_2 \ell_2} g_{\alpha_2 i_2,\beta_1 \ell_1}(t)]+m_{\alpha_2 i_2}[ m_{\beta_1 \ell_1} g_{\alpha_1 i_1,\beta_2 \ell_2}(t)+ m_{\beta_2 \ell_2} g_{\alpha_1 i_1,\beta_1 \ell_1}(t)]\\
    &+m_{\alpha_1 i_1} g_{\alpha_2 i_2,\beta_1 \ell_1 \beta_2 \ell_2}(t)
    +m_{\alpha_2 i_2} g_{\alpha_1 i_1,\beta_1 \ell_1 \beta_2 \ell_2}(t)
    +m_{\beta_1 \ell_1} g_{\alpha_1 i_1 \alpha_2 i_2, \beta_2 \ell_2}(t)
    +m_{ \beta_2 \ell_2} g_{\alpha_1 i_1 \alpha_2 i_2,\beta_1 \ell_1}(t)\\
    &+ g_{\alpha_1 i_1 \alpha_2 i_2,\beta_1 \ell_1\beta_2 \ell_2}(t)
    -\expval{[s_{\alpha_1 i_1}(t)-m_{\alpha_1 i_1}][s_{\alpha_2 i_2}(t)-m_{\alpha_2i_2}]} \expval{[s_{\beta_1 \ell_1}(0)-m_{\beta_1 \ell_1}][s_{\beta_2 \ell_2}(0)-m_{\beta_2 \ell_2}]}~.
    \end{aligned}
    \end{equation}
    \end{widetext}
    The correlation function presented above is expressed not only using second-order terms such as $g_{\alpha i,\beta \ell}(t)$, but also involves correlation functions pertaining to higher-order fluctuations. These include terms such as $g_{\alpha_1 i_1 \alpha_2 i_2,\beta \ell}(t)$, $g_{\alpha i,\beta_1 \ell_1 \beta_2 \ell_2}(t)$, and $g_{\alpha_1 i_1 \alpha_2 i_2,\beta_1 \ell_1 \beta_2 \ell_2}(t)$.
    As in the previous Section, we assume the applicability of the Wick's theorem to treat the correlation functions involving higher-order fluctuations.  In addition to Eq.~\eqref{app-eq:Wick1}, we assume the following relations
    \begin{equation}\label{app-eq:Wick3}
    \begin{aligned}
    g_{\alpha i,\beta_1 \ell_1 \beta_2 \ell_2}(t) &\to [\expval{s_{\beta_1 \ell_1}(0)}-m_{\beta_1 \ell_1}]g_{\alpha i,\beta_2 \ell_2}(t)\\
    &+[\expval{s_{\beta_2 \ell_2}(0)}-m_{\beta_2 \ell_2}]g_{\alpha i,\beta_2 \ell_2}(t)~,
    \end{aligned}
    \end{equation}
    and 
    \begin{widetext}
    \begin{equation}\label{app-eq:Wick4}
    \begin{aligned}
    g_{\alpha_1 i_1 \alpha_2 i_2,\beta_1 \ell_1 \beta_2 \ell_2}(t) &\to 
     g_{\alpha_1 i_1,\beta_1 \ell_1}(t)g_{\alpha_2 i_2,\beta_2 \ell_2}(t)
    + g_{\alpha_1 i_1,\beta_2 \ell_2}(t)g_{\alpha_2 i_2,\beta_1 \ell_1}(t)~\\
    &+\expval{[s_{\alpha_1 i_1}(t)-m_{\alpha_1 i_1}][s_{\alpha_2 i_2}(t)-m_{\alpha_2i_2}]} \expval{[s_{\beta_1 \ell_1}(0)-m_{\beta_1 \ell_1}][s_{\beta_2 \ell_2}(0)-m_{\beta_2 \ell_2}]}~.
    \end{aligned}
    \end{equation}
    \end{widetext}
    As discussed below Eq.~\eqref{app-eq:Wick1}, because we are examining the stationary limit, the expectation value is given by $\expval{s_{\alpha i}(t)}=m_{\alpha i}$. As a result, any term of the form $g_{\alpha_1 i_1 \alpha_2 i_2,\beta \ell}(t)$ and $g_{\alpha i,\beta_1 \ell_1 \beta_2 \ell_2}(t)$ vanishes.
    By imposing Eqs.~\eqref{app-eq:Wick1},~\eqref{app-eq:Wick3} and~\eqref{app-eq:Wick4} onto the generic correlation functions defined in Eq.~\eqref{app-eq:ss-ss}, one has
    \begin{widetext}
    \begin{equation}\label{app-eq:Wss}
    \begin{aligned}
    &\expval{[ s_{\alpha_1 i_1}(t) s_{\alpha_2 i_2}(t)-\expval{s_{\alpha_1 i_1}(t) s_{\alpha_2 i_2}(t)}][ s_{\beta_1 \ell_1}(t) s_{\beta_2\ell_2}(t)-\expval{s_{\beta_1 \ell_1}(t) s_{\beta_2 \ell_2}(t)}]}\to\\
    &\to m_{\alpha_1 i_1}[ m_{\beta_1 \ell_1} g_{\alpha_2 i_2,\beta_2 \ell_2}(t)+ m_{\beta_2 \ell_2} g_{\alpha_2 i_2,\beta_1 \ell_1}(t)]+m_{\alpha_2 i_2}[ m_{\beta_1 \ell_1} g_{\alpha_1 i_1,\beta_2 \ell_2}(t)+ m_{\beta_2 \ell_2} g_{\alpha_1 i_1,\beta_1 \ell_1}(t)]\\
    &+ g_{\alpha_1 i_1,\beta_1 \ell_1}(t)g_{\alpha_2 i_2,\beta_2 \ell_2}(t)
    + g_{\alpha_1 i_1,\beta_2 \ell_2}(t)g_{\alpha_2 i_2,\beta_1 \ell_1}(t) ~.
    \end{aligned}
    \end{equation}
Using the above result, one can express in closed form the two components of the resistance-resistance correlation functions, where the one associated with the out-of-plane magnetic fluctuations is
\begin{equation*}
\begin{aligned}
 \expval{\delta R_\parallel(t) \delta R_\parallel(0)}&=
(J+ \Delta)^4
\{B^2(\xi)[g_{Az,Az}(t)+g_{Bz,Bz}(t)-g_{Az,Bz}(t)-g_{Bz,Az}(t)]\\
&+g_{Az,Az}(t) g_{Bz,Bz}(t)+g_{Az,Bz}(t) g_{Bz,Az}(t)
\}~,
\end{aligned}
\end{equation*}
with $\sigma_z= \sqrt{1-B^2(\xi) - 2B(\xi)/\xi}$, and the second one associated with the in-plane magnetic fluctuations is represented as
\begin{equation*}
\begin{aligned}
\expval{\delta R_\bot(t)\delta R_\bot(0)}&=2\Big\{
(J^4+\Omega_0^4)[g^2_{Ax,Ax}(t)+g^2_{Ax,Bx}(t)]
-(J^4-\Omega_0^4)[g^2_{Ax,Ay}(t)+g^2_{Ax,By}(t)]\\
&-4 \Omega_0^2 J^2
g_{Ax,Ax}(t)g_{Ax,Bx}(t)\Big\},
\end{aligned}
\end{equation*}
\end{widetext}
where $\Omega_0=\sqrt{\Delta(2J+\Delta)}$. In agreement with the discussion  in  Appendix~\ref{ssect:in-plane}, we have assumed so far $t>0$. However, for $t<0$, due to stationarity and the commutativity of time dependence in the two-time correlation functions, it follows that $\expval{R_z(t)R_z(0)}=\expval{R_z(|t|)R_z(0)}$ and $\expval{R_\bot(t)R_\bot(0)}=\expval{R_\bot(|t|)R_\bot(0)}$.

\normalem 
\bibliography{bibliography}

\end{document}